\documentclass[prx,twocolumn,showpacs,longbibliography,nofootinbib, superscriptaddress]{revtex4-1}

\usepackage{graphicx}
\usepackage{amssymb}
\usepackage{amsmath}
\usepackage{esint}
\usepackage{epstopdf}
\usepackage{color}
\usepackage{ulem}

\newcommand{\ie}{{\it{i.e.}} }
\newcommand{\eg}{{\it{e.g.}}}

\newcommand{\be}{\begin{eqnarray}}
\newcommand{\ee}{\end{eqnarray}}
\newcommand{\bfig}{\begin{figure}}
\newcommand{\efig}{\end{figure}}

\newcommand{\rhoxy}{\rho_{xy}}

\newcommand{\rhoxyT}{\rho_{xy}^T}

\newcommand{\black}{\color{black}}
\newcommand{\red}{\color{red}}

\newcommand{\crgb} {CeRu$_2$Ga$_2$B}
\newcommand{\chiac} {\chi_{\rm ac}}
\newcommand{\dchidh} {\frac{d\chiac}{dH}}
\newcommand{\muh} {\mu_0H}
\newcommand{\upcirc} {^{\circ}}

\begin{document}

\title{Evolution of magnetic bubble domains in the uniaxial ferromaget \crgb~ inferred  from the Hall effect and ac magnetic susceptibility }
\author{Peter E. Siegfried$^{\dag}$}
\affiliation{Department of Physics, University of Colorado, Boulder, CO 80309, 
USA}%
\author{Mark Maus$^{\ddag}$}
\affiliation{Department of Physics, University of Colorado, Boulder, CO 80309, 
USA}%
\author{Alexander C. Bornstein}
\affiliation{Department of Physics, University of Colorado, Boulder, CO 80309, 
USA}%
\author{Dirk Wulferding}
\affiliation{Department of Physics and Astronomy, Sejong University, Seoul, South Korea}%
\author{Jeehoon Kim }
\affiliation{Department of Physics, POSTECH,  Pohang, South Korea}%
\author{Ryan E. Baumbach}
\affiliation{Physics Department, University of California Santa Cruz, Santa Cruz, CA 95064, USA}%
\author{Eric D. Bauer}
\affiliation{Los Alamos National Laboratory, Los Alamos, New Mexico 87545, USA}%
\author{Filip Ronning}
\affiliation{Los Alamos National Laboratory, Los Alamos, New Mexico 87545, USA}%
\author{Minhyea Lee}
\affiliation{Department of Physics, University of Colorado, Boulder, CO 80309, 
USA}%
\email {minhyea.lee@colorado.edu}

\date{\today}

\begin{abstract}
We study  the Hall effect,  AC magnetic susceptibility ($\chiac$), and magnetic force microscopy of the uniaxial ferromagnet \crgb~ with a centrosymmetric crystal structure. 
 We observe a finite topological  Hall effect (THE)  within the ordered phase, before the magnetization is polarized by applied field. By comparing the field dependences of the area fraction of the magnetic bubbles, the derivative of $\chiac$, and the THE signal, we deduce that the magnetic bubbles in \crgb~ evolve from the trivial to topological spin texture with field.
 Our findings enable the expansion of the search for magnetic materials hosting topological spin textures to include uniaxial ferromagnets and open a new possibility to tailor the topological spin texture.\black
  \end{abstract}

\maketitle


\section{Introduction}

Intermetallic $f$-magnetic materials have long been of central interest
 as they host a wide range of fascinating magnetic phases,
including  complex local moment order \cite{Fushiya2014, Sunku2016, LYe2017} , non-Fermi liquid behavior \cite{Lohneysen2007_RMP},
hidden  ordered states \cite{Mydosh2011} and  superconductivity adjacent to the magnetic phases  \cite{Pfleiderer2009_RMP}.  
The complex interplay between the lattice, charge, orbital,
and spin degrees of freedom often are mediated by  a variety of  interaction channels such as Ruderman-Kittel-Kasuya-Yoshida (RKKY) interaction \cite{Ruderman1954, Kasuya1956, Yoshida1957},  crystal electric field splitting \cite{Newman1971, Shin2020},  and the Kondo interaction \cite{Kondo1964,Terashima2002, Seiro2018}. 
 
Recently, there has been a surge of interest in a new group of centrosymmetric lanthanide intermetallics that host  Skyrmion states  \cite {Lin2018, Kurumajii2019, Hirschberger2019} 
or non-trivial spin texture inferred from other physical quantities \cite{Leahy2022, Fruhling2024}. 
In these materials, instead of a Dzyaloshinskii-Moriya interaction (DMI),   competition among strong anisotropy, 
magnetic dipole-dipole interaction and Zeeman energy play a crucial role in stabilizing non-trivial spin texture \cite{Ezawa2010, Kiselev2011}. 
This is contrasted with extensive studies of the non-centrosymmetric magnets such as MnSi \cite{Muhlbauer2009},  Fe$_{0.5}$Co$_{0.5}$Si \cite{XZYu2010}, FeGe,\cite{XZYu2011}, and CeAlGe \cite{Piva2023},  where a competition between ferromagnetism and DMI enables non-collinear and non-coplanar spin textures like the Skyrmion lattice \cite{Rossler2006,Nagaosa2013}. 

A  circular-shaped region in which the enclosed spins point in different directions from the surrounding area is referred to as a magnetic bubble.  This includes magnetic vortices and domain walls and has been a familiar concept in nano- and micro-scale magnetism for many decades  \cite{Leeuw1980}.    The formation of bubble-like domains is commonly observed in strong uniaxial magnets as it balances the magnetic dipole interaction among domains and anisotropy. 
The discovery of a magnetic Skyrmion lattice phase in MnSi \cite{Muhlbauer2009} renewed great interest in these topological defects in magnetic systems thanks to their spectacular dynamical properties and emergent electromagnetic phenomena that offer great potential for novel magnetic storage and other spintronics applications \cite{Nagaosa2013,YFLi2013, Fert2013}. 

\red
\black

 \crgb~is a centrosymmetric metallic ferromagnet with a Curie temperature $T_C = 15.5$ K, 
 where the local moments from Ce $4f$-electrons order without significant hybridization with the conduction electron state \cite{Baumbach2012JoP}.
 It exhibits highly uniaxial magnetic anisotropy with the magnetic easy axis along the $c$-axis, where a small applied field ($H_S =$  0.15 T) brings the system to the spin-polarized state, while the same magnitude of field applied in the $ab$ plane barely induces any magnetization [Fig. \ref{MagMR}(a)]. 
 The strong Ising-like anisotropy is thought to originate from the crystal electric field effect of Ce$^{3+}$  single ion anisotropy \cite{Sakai2012, Matsuoka2012}. 

Previously, magnetic bubble domains on the $ab$-plane of \crgb~  were  observed by Wulferding {\it{et al.}} \cite{Wulferding2017}, via magnetic force microscopy (MFM) under applied field along the $c$-axis below $T_C$ .
The bubble-like domains are observed unambiguously in MFM due to its superb sensitivity to changes in the out-of-plane components of magnetization: however, the limited spatial resolution of the MFM makes it hard to reveal the local spin configuration within a bubble to determine so-called topological charge ($\mathcal N$), which refers to the number that counts how many times the arrows of spin directions within  the structure wrap around a sphere \cite{Rossler2006,Nagaosa2013}: 
 \be
\mathcal N = \frac{1}{4\pi} \oiint_{\mathcal A} \mathbf n\cdot \Big(\frac{\partial \mathbf n}{\partial x}\times\frac{\partial \mathbf n}{\partial y}\Big)~dxdy,
\label{topocharge}
\ee 
where $\mathcal A$ is the area of a bubble, and $\mathbf n(\mathbf r)$ denotes the unit vector in the direction of spins at a position $\mathbf r$.  


If a spin texture extended to macroscopic length scales contains these bubbles with non-zero topological charge (\eg~ $\mathcal N=1$ for a skyrmion), it should lead to an associated macroscopic electromagnetic response  \cite{Kiselev2011, Nagaosa2013, Hayami2018}.
These include but not limited to the topological Hall effect (THE) \cite{Lee2009,Neubauer2009,Schulz2012},  magnetic AC susceptibility \cite{Bauer2012, Bauer2017}, specific heat \cite{ Bauer2013}  and magnetoresistance   \cite {Siegfried2017, Lobanova2016} similar to what was observed in the skyrmion lattice phase.

In this paper, we present the field ($H$) and temperature ($T$) dependence of the topological Hall resistivity ($\rhoxyT$) as well as ac magnetic susceptibility ($\chiac$) of the ordered phase of \crgb~($T<T_C$). We trace the variation of the topological charge using $\rhoxyT$ as a gauge by  comparing it  to the area fraction change of bubble domains observed in MFM 
as a function of field. 
At a fixed $T$, $\rhoxy^T$  reaches its maximum at  $H  = H_T$, which is narrowly apart from another field scale $H_P$, where the derivatives of $\chiac$  with respect to field   ($\dchidh$)   exhibit the maximum magnitude. This is  similar to what was observed in B20 magnet MnSi, where 
 $\chiac (H)$ exhibits a step-like change and $\dchidh$ a delta function \cite{Bauer2012, Wilhelm2011,Bannenberg2018} upon the formation of the Skyrmion Lattice. Meanwhile, such changes in \crgb~appear with considerable broadening.  
 We also find that the bubble area fraction remains constant for $H<H_T$, but decreases rapidly as $ H$ is increased further, in the same manner as the THE signal. 
 
 Based on our observation, we speculate the magnetic bubble in \crgb~  undergoes the following evolution with field: it begins as a trivial magnetic domain with zero topological charge ($i.e.$ collinear or coplanar or both)  in zero field. 
 As the field increases, the topological charge within a  bubble begins to develop locally and results in a finite   $\rhoxyT$ signal.  As the field further increases, $\rhoxyT$ diminishes to zero as the field approaches the saturation field.  
  Furthermore, based on the analogy to the behavior of $\dchidh$ in B20 magnet \cite{Bauer2012, Wilhelm2011, Bannenberg2018}, the broad peak in $\dchidh(H)$ indicates the wide variation of the onset field values for the topological charge to emerge in individual magnetic bubbles in \crgb. 

Our findings will be utilized to expand the search for magnetic materials hosting the topological spin textures to ones with uniaxial anisotropy,  and open a new avenue to tailoring topological defects in magnetic thin films, of which the coplanar magnetic bubbles are ubiquitously observed, some of which turns out  topological ones  \cite{Vistoli2019, Nakamura2018, Maccariello2018, Jeong2015}.

 \red
%
%
%
\black

\section {Experimental Methods}

Single crystals of \crgb~were grown by the Czochralski method \cite{Baumbach2012JoP}.
Magnetization ($M$) data were measured using a Quantum Design magnetic properties measurement system. 
Magneto transport measurements were performed in a standard Hall-bar geometry, where the contact was made using Dupont Ag paints with a typical contact resistance of $\le 1~\Omega$.  
The sample dimensions range (1--2) mm $\times$(0.5--1) mm with varying  thickness  along the $c$-axis. For the Hall measurements, it was 0.4 and 0.6 mm  and for the AC susceptibility, 0.9 -- 1.2 mm. \black
All the magneto transport data are obtained in the applied magnetic field along the crystalline $c$-axis and the  DC current of  2 -- 6 mA was flown within the $ab$-plane.  The magnetic field was generated by the  7 T superconducting magnet sitting in the liquid Helium bath.

The ac magnetic susceptibility measurements were performed with a home-built nesting coil set comprising a driving coil to generate a very small ac magnetic field and a pair of pick-up coils, coaxially placed within the driving coil  \cite{Maus2019}. The two pick-up coils are wound in opposite directions and connected in series to ensure equal mutual inductance with opposite sign. This results in zero net induced voltage in a null state. 
When the sample is placed in only one of the pick-up coils, a nonzero net voltage is induced, the magnitude of which is directly proportional to the change of magnetic flux contributed by the sample.  This voltage is measured with a digital lock-in at a driving frequency and amplitude of 997 Hz and 7.1 gauss, respectively. 
Heat sinking for the sample inside the coil was achieved by adhering the samples to the long, thin sapphire rod using GE varnish. This rod is then inserted through the pick-up coils and anchored to the cold finger. The crystalline $c$-axis is aligned to the co-axis of the pick-up and driving coils that are parallel to the applied DC field. We find the background signal of coils is negligible compared to the signal generated by the presence of the samples.  

MFM images were taken with a home-built MFM probe inside a cryostat with superconducting magnets. Details of the setup are described elsewhere \cite{Jeong2015, Yang2016, Wulferding2017}. 

For the rest of this article, the applied magnetic field direction is always parallel to the crystalline $c$-axis, $\mathbf H\parallel c$, unless otherwise specified.

\section{Results and Discussion}

\begin{figure}[ht]
\begin{center}
\includegraphics [width=\linewidth]{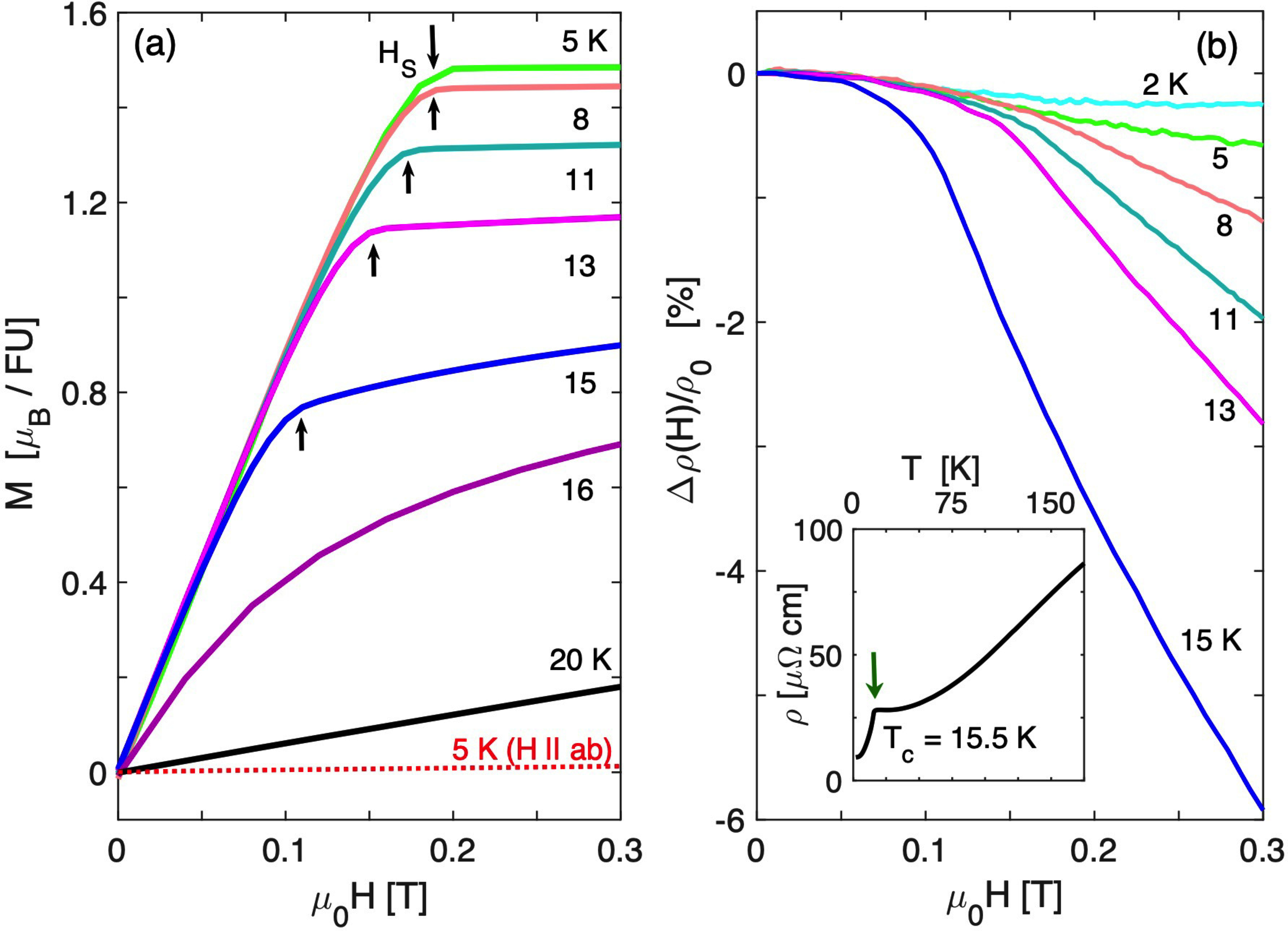}
\caption{\small   (a) Magnetization as a function of applied magnetic field $H$ in $\mathbf H\parallel c$ at different temperatures. The saturation of $M(H)$ at $H_S$ is indicated for each temperature (black arrow) and $H_S$ decreases as $T$ approaches $T_C=15.5$ K. The magnitude of $M$ for $\mathbf H\parallel ab$ is minuscule (dotted line), compared to $\mathbf H\parallel c$. (b) The fractional MR at various temperatures is shown as a function of $H$. Below $T_C$, MR changes little in $H<H_S$. The inset shows the zero field resistivity, where $T_C$ is clearly marked (arrow) by the rapid drop in resistivity with decreasing temperature.}
\label{MagMR}
\end{center}
\end{figure}

Fig.~\ref{MagMR}(a) displays the magnetization ($M$) of \crgb~ as a function of field along the $c$-axis at various temperatures. A saturated moment of 1.55$\mu_B$ per Ce is reached at $\muh_S \simeq 0.18$ T and remains constant up to $\muh = 7$ T, at $T=5$ K.
The saturation field $H_S$, marked by arrows, decreases as $T$ gets close to $T_C$. 
The linear increase of $M_{C}$ up to the finite $H_S$  indicates the slow polarization of abundant magnetic domains in the zero field (ZF) state.
We find that the magnetization in  $\mathbf H \parallel ab$  is two orders of magnitude smaller than $M$,  which highlights the strong uniaxial magnetic anisotropy of the system.

Fig.~\ref{MagMR}(b) shows the fractional MR (fMR), defined as 
$\Delta\rho(H)/\rho_0 = [\rho(H)-\rho_0]/\rho_0$, where $\rho$ refers to the longitudinal resistivity, and  $\rho_0= \rho(H=0)$ at a given $T$. 
The inset shows $\rho$  vs $T$ at ZF, where $T_C$ is marked by a sharp drop  in $\rho(T)$ (green arrow).  
$M_c(H)$ and fMR in the extended field range up to 1 T are shown in Fig. 7 in the Appendix. \black

The fMR of \crgb~is found negative in $T<T_C$, and its magnitude ranges a few \% in $H<H_S$ below $T_C$. 
Changes in MR for $H>H_S$ become more obvious once $T$ increases close to $T_C$,  due to the effective reduction of spin scattering with the conduction electrons in the fully polarized local spin state in $H >H_S$. 
A sharp drop in the $\rho$ vs $T$ plot, shown in the inset and marked by a green arrow, confirms the reduced spin scattering upon ordering.
Large changes of resistivities in the vicinity of  $T_C$ with applied magnetic fields are  a well-known feature of metallic ferromagnets across both $d$- and $f$-electron systems \cite{Campbell1982}. 
 Near $T_C$, short-range spin fluctuations strongly enhance 
 scatterings of conduction electron; applying a magnetic field suppresses these fluctuations, producing a rapid decrease in resistivity and consequently a large (negative) magnetoresistance.

 This phenomenon had been recognized since late 1950's since the seminal studies  shown in Ref. \cite{ deGennes1958, Yosida1957, Fisher1968, Alexander1976}, where the connection between the electrical resistivity and critical behavior near $T_C$  is established. Because large magnetoresistance is technologically attractive for sensor applications and spin-fluctuation suppression is most efficient in this critical regime, the effect remains an active research topic?especially in ferromagnets with near room temperature, e.g., Refs. \cite{MWang2021,Chakravorty2015}. 
\black

Next,  we show the Hall resistivity ($\rhoxy$)  of \crgb~as a function of applied field in Fig.~\ref{Hall}(a) 
at $T =2, 5, 8,11,13, 15$ and 20 K. Similar to the $M(H)$, the saturation field $H_S$ is clearly identifiable by the kink in $\rhoxy(H)$.

The ordinary Hall coefficient, $R_H$ at a given $T$ is obtained from the linear high-field slopes 
in $H>H_S$ up to 1T  for all $T$'s  here as shown  Fig. 7 in Appendix. \black  
$R_H$ is negative for all $T$, except for $T=20$ K at which $\rhoxy (H)$ does not show the linear $H$ dependence up to 7 T. The positive slope of $\rhoxy$ at 20 K simply reflects the anomalous Hall signal ($\rhoxy^A$) induced by field-induced magnetization,  where  the kink marking $H_S$ below  $T_C$ has moved out to the higher field, leaving the AHE with the positive slope. This is commonly observed in ferromagnetic intermetallic systems \cite{Checkelsky2008}.

\bfig[b!]
\begin{center}
\includegraphics[width=\linewidth]{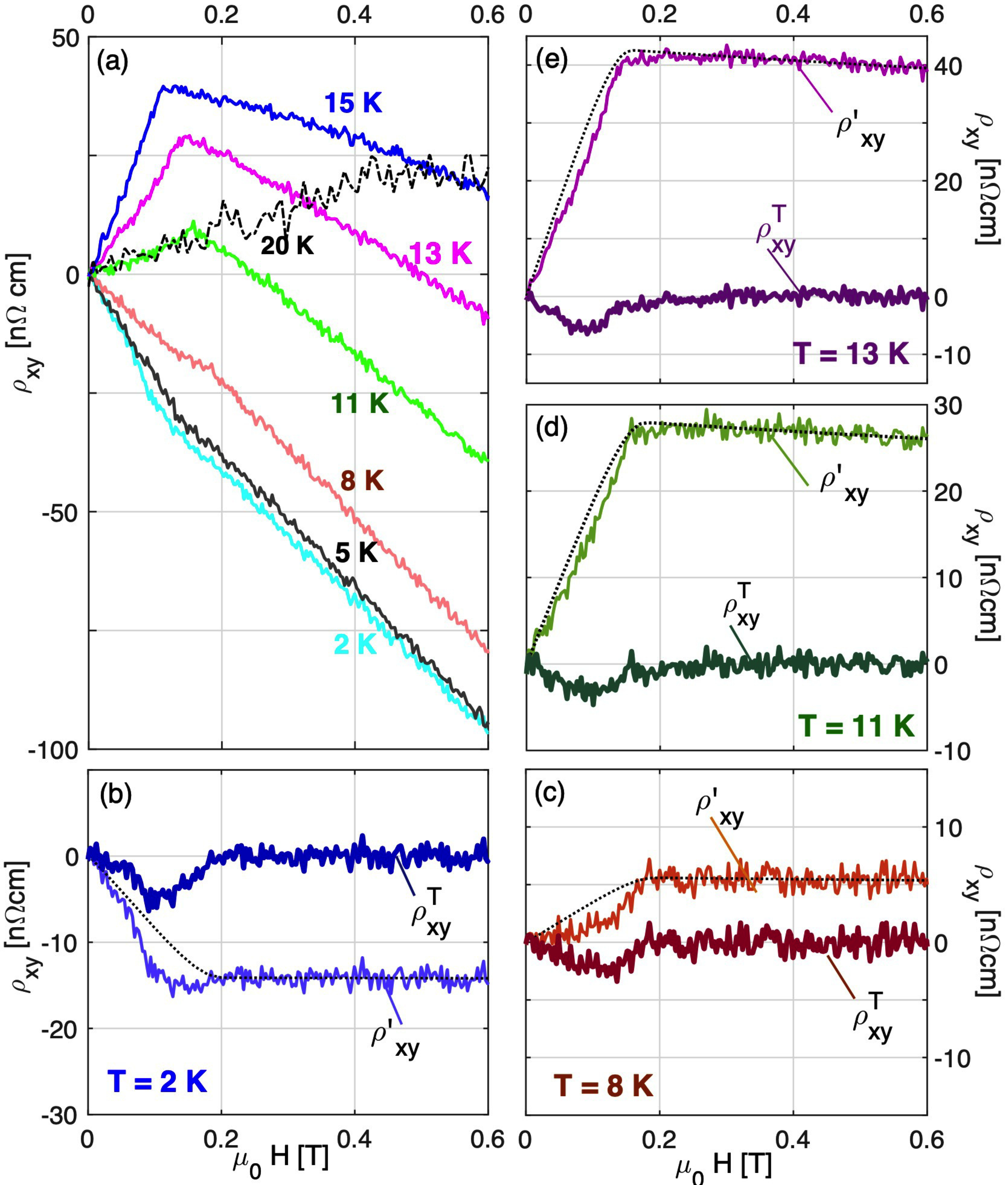}
\caption{\small (a) $\rhoxy(H)$ measured at $T= 2, 5, 8, 11, 13, 15$ and 20 K.
(b-e) Separating out the THE ($\rhoxy^T$) from $\rhoxy' = \rhoxy - \mu_0R_HH$ for 
(b) $T = 2$ K, (c) 8 K, (d) 11 K, and (e) 13 K. $\rhoxy^T(H)$ (darker, bold lines) is determined by subtracting $\rhoxy^A(H) = S_H\rho^2M$ (dotted lines) from $\rhoxy^{'}$ in each panel. See text for details.}
\label{Hall}
\end{center}
\efig

In order to identify the different contributions to the Hall resistivity in $T<T_C$, 
first, we parse the $\rhoxy(H)$ into the ordinary Hall contribution that is linear in $H$ at high field ($\mu_0R_HH$) and the remainder of signal, $\rhoxy^{'}$ \cite{Lee2007}.
$\rhoxy^{'}$ includes both the anomalous Hall resistivity $\rhoxy^A$, and the topological Hall effect (THE) contribution $\rhoxy^T$. 
The former is proportional to the overall magnetization,  and the latter is attributed to  the topological charge arising from non-spin texture within the magnetic bubbles.
\begin{eqnarray}
\rhoxy^{'} (H) &=& \rhoxy(H)-\mu_0R_HH = \rhoxy^A + \rhoxy^T
\label{rxyeq}
 \end{eqnarray}
When the $H$-linear ordinary Hall contribution ($\mu_0R_HH$) is subtracted from $\rhoxy(H)$ to obtain $\rhoxy^{'}$ in \crgb, we find that  $\rhoxy^{'}$ cannot be captured by the anomalous Hall effect  (AHE)  alone: The anomalous Hall resistivity ($\rhoxy^A$) is proportional to $M$ with the anomalous Hall coefficient $R_S$ that is defined as  $\rhoxy^A = R_SM$, with $R_S$ being a function of both $H$ and $T$.  As $T$ increases, there is a change in the kink angle at $H_S$ between $T=5$ and 8 K [Fig. \ref{Hall} (a)], indicating the sign change of $R_S$. 

We find at a given $T$  $\rhoxy^A$ of \crgb~ is  best described  by $\rhoxy^A = S_H\rho^2M$, where   $R_S$ is expressed with  a $H$-independent constant $S_H$  and  $\rho(H)$, magneoresistance as shown in Fig. \ref{MagMR}(b)  \cite{Lee2007}.  It particularly captures better  the subtle concave curvature of $\rhoxy^{'}$ in  $H>H_S$  than using $R_S =  A\rho$, with $A$ is a proportional constant. Dotted lines in panels Fig.~\ref{Hall} (b-e) display the best-fit results of $\rhoxy^A = S_H\rho^2M$, at $T= 2, 8, 11$ and 13 K.

After carefully characterizing the anomalous Hall contribution, we can finally extract the topological Hall contribution from the differences between  $\rhoxy^{'}$ and $\rhoxy^A$
according to Eq. \ref{rxyeq} \cite{Lee2009, Neubauer2009, Huang2012}. 
$\rhoxy^T (H)$  of \crgb~ at each $T$'s is shown in the thicker lines in the panels (b-e). 
The finite $\rhoxy^T$ values exhibit similar $H$-profiles at all temperatures:  it starts from zero to increase  smoothly with $H$ and reaches its maximum value at $H=H_T$ before decreasing back to zero as $H \rightarrow H_S$. 
We infer from this non-monontonic behavior of $\rhoxy^T$  that \crgb~ hosts  the non-trivial spin texture that enables the finite THE signal upon applying magnetic field, but the onset field of such a texture may be broadly distributed across a range of field as   $\rhoxy^T$ rises smoothly with field. We note that the disappearance rate of $\rhoxy^T$ near $H_S$ is faster than the appearance rate near  ZF throughout all temperatures, indicating that the onset field of transition into the topological spin textures may have a larger distribution than the spin polarization field scale.  We will come back and discuss this further in Fig. \ref{chiac}.

While $\rhoxy^A(H)$ mostly follows the $H$-profile of the magnetization -- monotonically increasing with $H$ and then saturating, the sign change in $S_H$ between 5 and 8 K is noticeable by flipping $\rhoxy^{'}$ with respect to the horizontal axis. 
Meanwhile,  the field dependence of $\rhoxy^T$ exhibits a clear maximum at $H=H_T$, which varies slightly with $T$ and the sign remains negative throughout. 
Table 1 lists $R_H$ and $S_H$ and the maximum value of $\rhoxy^T$ at each temperature. 

\bfig[ht]
\begin{center}
\includegraphics[width=0.9\linewidth]{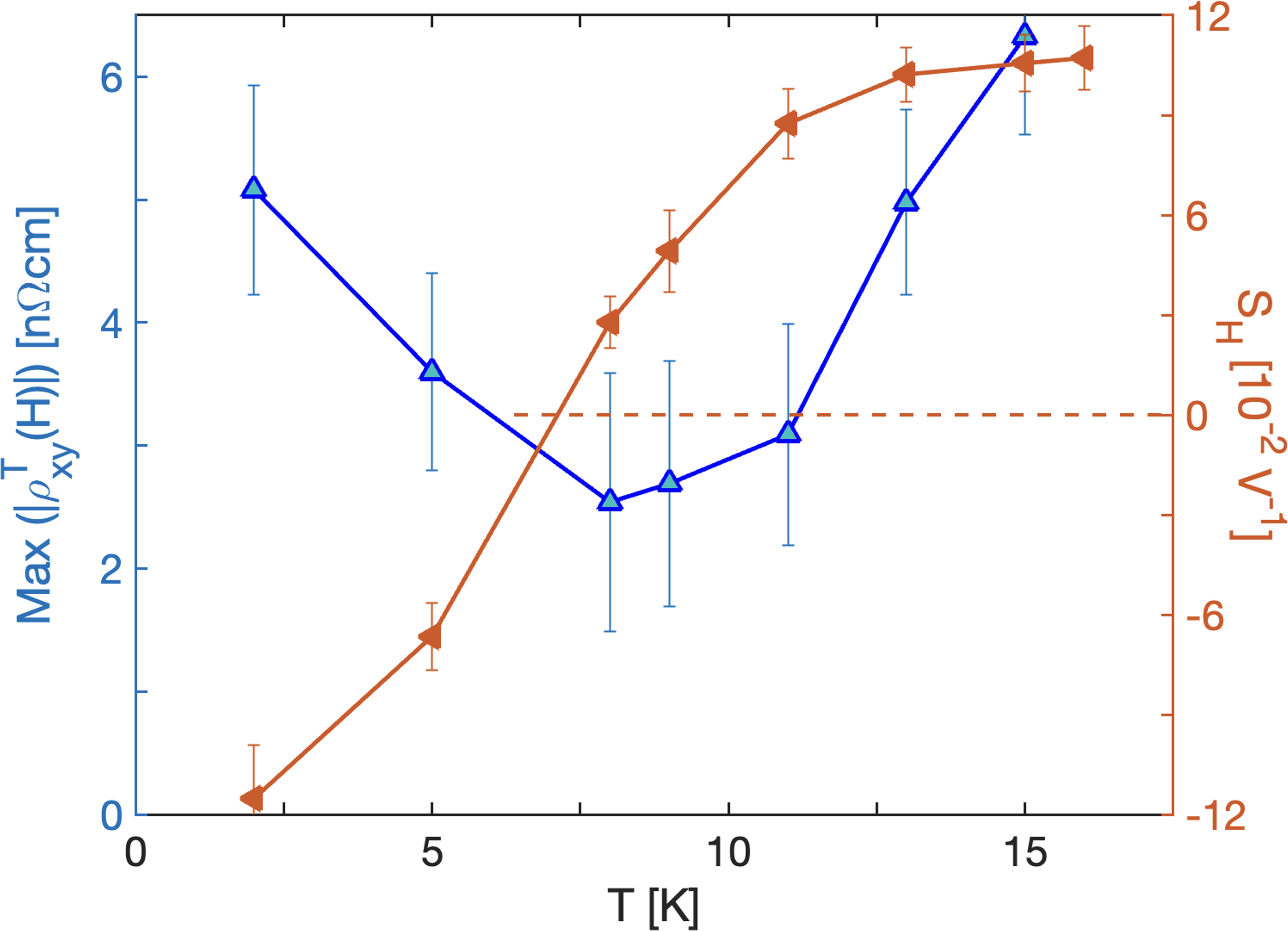}
\caption{\small Temperature dependence of the maximum magnitude of $|\rhoxy^T(H)|$ (left $y$-axis) and $S_H$ (right $y$-axis). The sign of $S_H$ changes from  negative to positive between 5 and 8 K, while that of $\rhoxy^T$  remains negative.
}
\label{SHmaxTHE}
\end{center}
\efig
 %
 
In Fig.~\ref{SHmaxTHE}, we plot $S_H$ (right $y$-axis) and the maximum value of $|\rhoxy^T|$ (left $y$-axis) as a function of $T$. 
The sign of $S_H$ changes from negative to positive between 5 and 8 K, while that of $\rhoxy^T$  remains negative.  
It is interesting to compare  the  sign change of $S_H$ to that of  Fe-doped MnSi samples.  In Mn$_{1-x}$Fe$_x$Si, it was found that the sign of the THE  remains the same \cite{Chapman2013}, while that of   $S_H$   depends on the Fe doping level.  
This is expected because the origin of the sign of AHE is more complex, involving material-specific traits such as the electronic wave functions leading to the Berry curvature, spin-orbit coupling and the level of disorders \cite{AHEreview2010}.
On the other hand,  the THE sign is set by the real space spin texture, e.g., the handedness of skyrmions and hence remains the same as long as the normal Hall coefficient remains the same  \cite{Kanazawa2011,Shiomi2012, Chapman2013, Franz2014}.

\begin{table}[b]
\centering
\caption{\small Fitting parameters used in parsing the different contributions of the Hall signal, $\rhoxy (H) = \mu_0R_HH + S_H\rho^2M + \rhoxy^T$, where $R_H$ and $S_H$ are $H$-independent constants at a given $T$.}
\begin{tabular}{cccc}
\hline\hline
$T$ & $R_{\rm H}$ & $S_H$ & {\rm max} $|\rho^{T}_{xy}|$ \\

 [K] & [10$^{-9}~$m$^3$/C ]               &$[ 10^{-2} ~\rm V^{-1} ]$     & [ n$\Omega$ cm ]       \\ 
\hline
2                    & -1.39                                      &$ -11.5\pm1.6 $                                          & $5.1\pm 0.8$     \\ 
5                    & -1.44                                  & $-6.7\pm1.0$                                                    & $3.6 \pm 0.8$      \\ 
8                    & -1.41                                     & $2.8 \pm 0.8$                                                     & $2.5 \pm 1.1 $ \\ 
9                    & -1.33                                      & $4.9  \pm 1.2 $                                                   & $2.7\pm1.0 $    \\ 
11                   & -1.08                                     & $8.8 \pm  1.0  $                                                 & $3.1 \pm 0.9$  \\ 
13                   & -0.79                                      & $10.2 \pm 0.8 $                                                  & $5.0\pm0.7 $   \\ 
15                   & -0.50                                      & $10.6 \pm 0.8 $                                                  & $6.3 \pm0.8$    \\ 
16                   & -0.57                                      & $10.7 \pm 0.9   $                                              & N/A      \\ 
\hline\hline
\end{tabular}
\label{FitVals}
\end{table}

The finite THE signal with non-monotonic field dependence below $H_S$ indicates that \crgb~ hosts a non-trivial (i.e. non-collinear and non-coplanar)  spin texture that supports the topological charge as expressed in Eq. (\ref{topocharge}). 
Hence, we hypothesize that the topological Hall contribution in \crgb~should originate from the magnetic bubbles observed in MFM images \cite{Wulferding2017} and that the spin texture within a bubble evolves with increasing field from zero. 

To further examine this hypothesis,  we investigate the magnetic AC susceptibility ($\chiac$). The field dependences of $\chiac$ exhibit extraordinary sensitivity upon the transitions to the Skyrmion lattice\, playing a key role in determining the phase boundary in the B20-type cubic magnets \cite{Bauer2012,Bannenberg2018, Wilhelm2011}.
 
Fig.~\ref{chiac}(a) displays the $H$ dependence of $\chiac$.  
Despite appearing to have a linear relation between  $M$  and $H$ in Fig.~\ref{MagMR}(a), 
$\chiac (H)$ ($\propto \frac{dM}{dH}$) reveals a significant $H$ dependence, 
far from constant, at all measured temperatures. 
Because $\chiac$ is highly sensitive to  changes in spin stiffness, this behavior   already indicates that  tthe field-induced realignment of the spin texture within domains proceeds in a more complex manner than a simple spin polarization.
\black


\bfig[ht]
\begin{center}
\includegraphics[width=\linewidth]{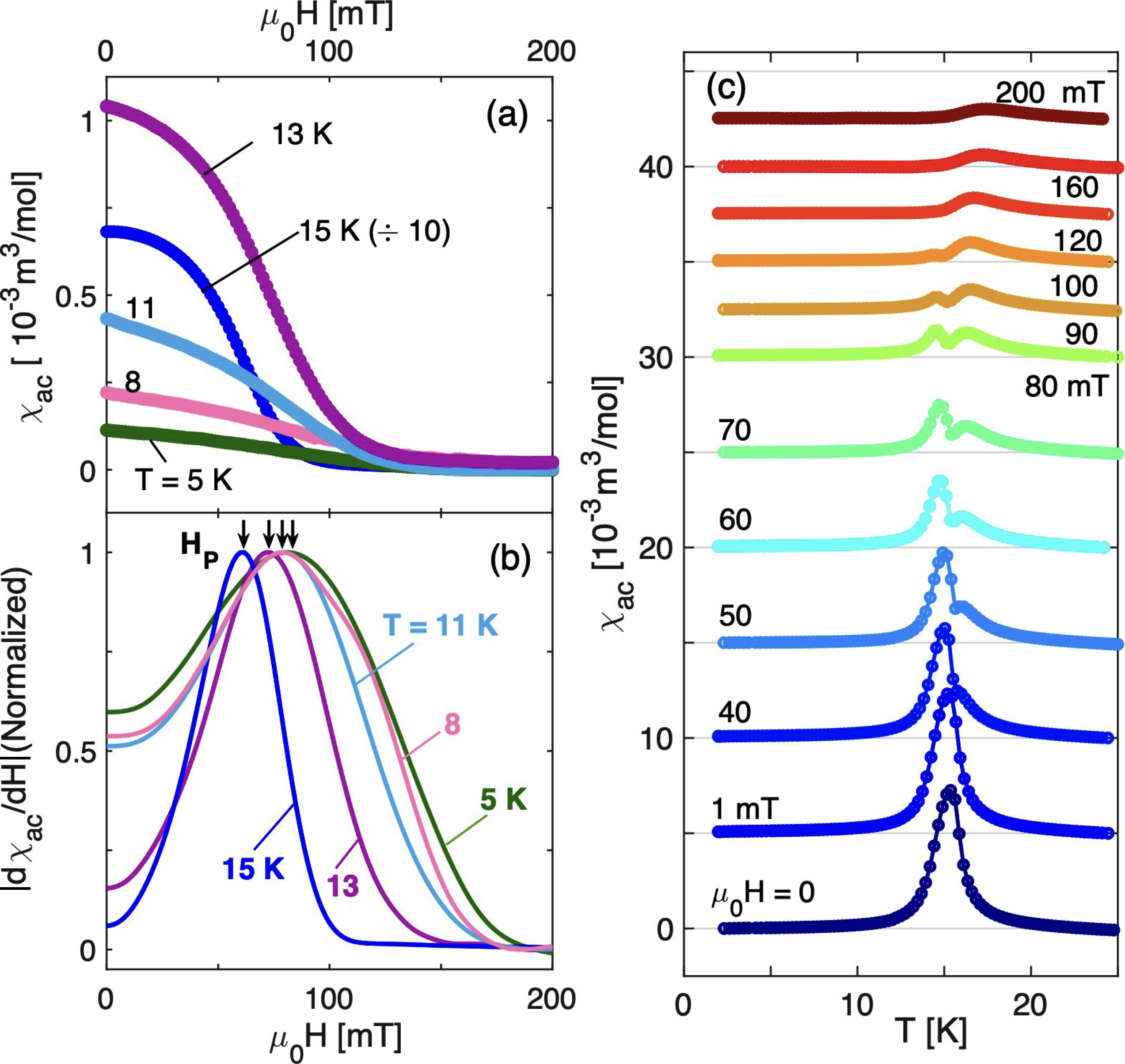}
\caption{\small (a)$\chiac$ plotted as a function of $H$. 
Note that $\chiac (H)$ at $T=15$ K is scaled by 1/10 in order to fit within the same range. 
(b) Derivative of $\chiac$ as a function of $H$, normalized by the maximum value at each $T$. Vertical 
arrows mark the location of the peak in $\dchidh$, $H_P$.
(c) $T$ dependence of $\chiac$ at fixed $H$. Each curve is displayed with an offset for clarity. 
} 
\label{chiac}
\end{center}
\efig

To understand better, 
 we take the derivatives of $\chiac(H)$ with respect to $H$, $\dchidh$: Fig. \ref{chiac} (b) displays the normalized  $\dchidh$ with the maximum  as a function of $H$ at each $T$'s.
In $H=0$ (ZF),  $|\dchidh|$ starts at a finite value in $T<11$ K, which decreases to zero in higher temperatures. 
At all temperatures, it exhibits a broad peak with its maximum occurring at  $H=H_P$ as marked with arrows. $H_P$ lies distinctly lower than  $H_S$'s, where the magnetic moment saturates, and it changes slowly with $T$, except  $T$ approaches $T_C$. The $T$ dependence of $H_P$ with other characteristic field scales of \crgb is shown in  Fig. \ref{phasedia}, which we discuss shortly. 

Here we point out that the overall  $\chiac(H)$ behavior of \crgb~  has substantial similarity to  that of B20-type cubic helimagnets, 
such as MnSi  \cite{Bauer2012}, Fe-doped MnSi  \cite{Bauer2010, Bannenberg2018}, and FeGe \cite{Wilhelm2011}:  in B20 magnets, entering the Skyrmion lattice phase with increasing field is unambiguously marked by almost step-like  changes in  $\chiac$ vs $H$ and the corresponding sharp peak/dip in $\dchidh$ at a given $T$.
Because a skyrmion lattice in these B20 helimagnets forms a high symmetry long-range pattern, it is not surprising to leave explicit thermodynamic signatures in $\chiac$ as well as specific heat \cite{Bauer2013} to mark the transition field from the trivial to topological spin textures.   
Unlike these B20 magnets,  both signatures in $\chiac (H) $ and $\dchidh$ in \crgb are found  dampened,  which is similar to what was observed in the B20 magnets with abundant pinning centers  \cite{Bauer2010, Bannenberg2018}\black. 

With these observations, we hypothesize that the  transformation 
from trivial to topological spin textures driven by applied field is a crossover rather than a well-defined phase transition, and that the individually localized bubbles undergo this evolution at different field scales. 
Hence,   a fraction of the magnetic bubbles  observed even at zero field \cite{Wulferding2017} 
should have a coplanar and non-collinear magnetic vortice-like texture that exhibit zero topological charge [Eq. (1)]. 
With increasing field, it acquires out-of-plane components aided by the uniaxial anisotropy to have finite topological charges, leading to non-zero $\rhoxyT$, as shown in  Fig. 2. 
This transformation is captured in  $\chiac(H)$ of \crgb~ by non-zero $\chiac$ 
at ZF  and  a smooth decrease in $\chiac(H)$ at $H_P$, which is far lower than $H_S$. The latter is more obvious in $|\dchidh|$ vs $H$ with a maximum at $H_P$.  In analogy with the behaviors of $\chiac$ and combined with the finite topological Hall signal,  these features clearly point to the changes of the topological nature of the magnetic bubbles in \crgb, yet they  are more broadened than what is observed in B20 magnets. \black

\begin{figure}[t]
\begin{center}
\includegraphics[width=\linewidth]{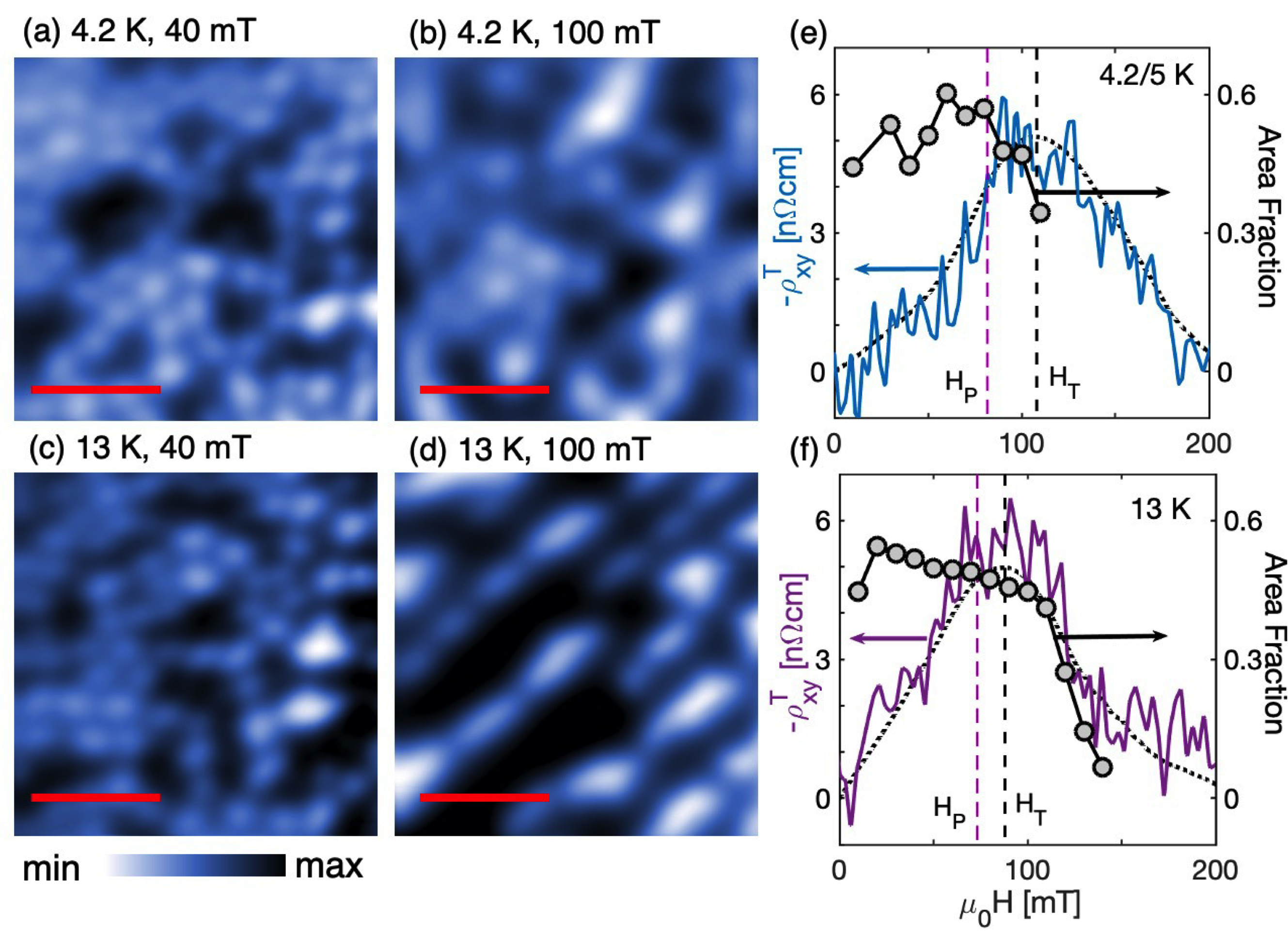}
\caption{\small MFM images at $\muh=40$ (left) and 100 mT (right)   at $T= 4.2$ K(a,b)  and at $T=13$ K  (c,d) with the scan area  $14\times14$ $\mu$m$^2$. All images were taken  with initially field-cooled at $\muh=10$ mT.  Scale bar corresponds to 5 $\mu$m.
The colorbar indicates the linear scale between the minimum and maximum frequency shifts of the cantilever at each $T$.
(e) $\rhoxy^T (H)$ (solid line, left $y$-axis) at $T=5$ K and area fraction of the bubble-like domains (gray circles, right $y$-axis) at $T=4.2$ K as a function of $H$  and (f) at 13 K. 
Dotted lines are a guide for the eyes for $\rhoxy^T$.  $H_P$ and  $H_T$  are marked with the vertical dashed lines, where the maximum of $\dchidh$ and $\rhoxy^T (H)$  are found, respectively. 
}
\label{mfmdata}
\end{center} 
\end{figure}
\begin{figure}[h]
\begin{center}
\includegraphics[width=\linewidth]{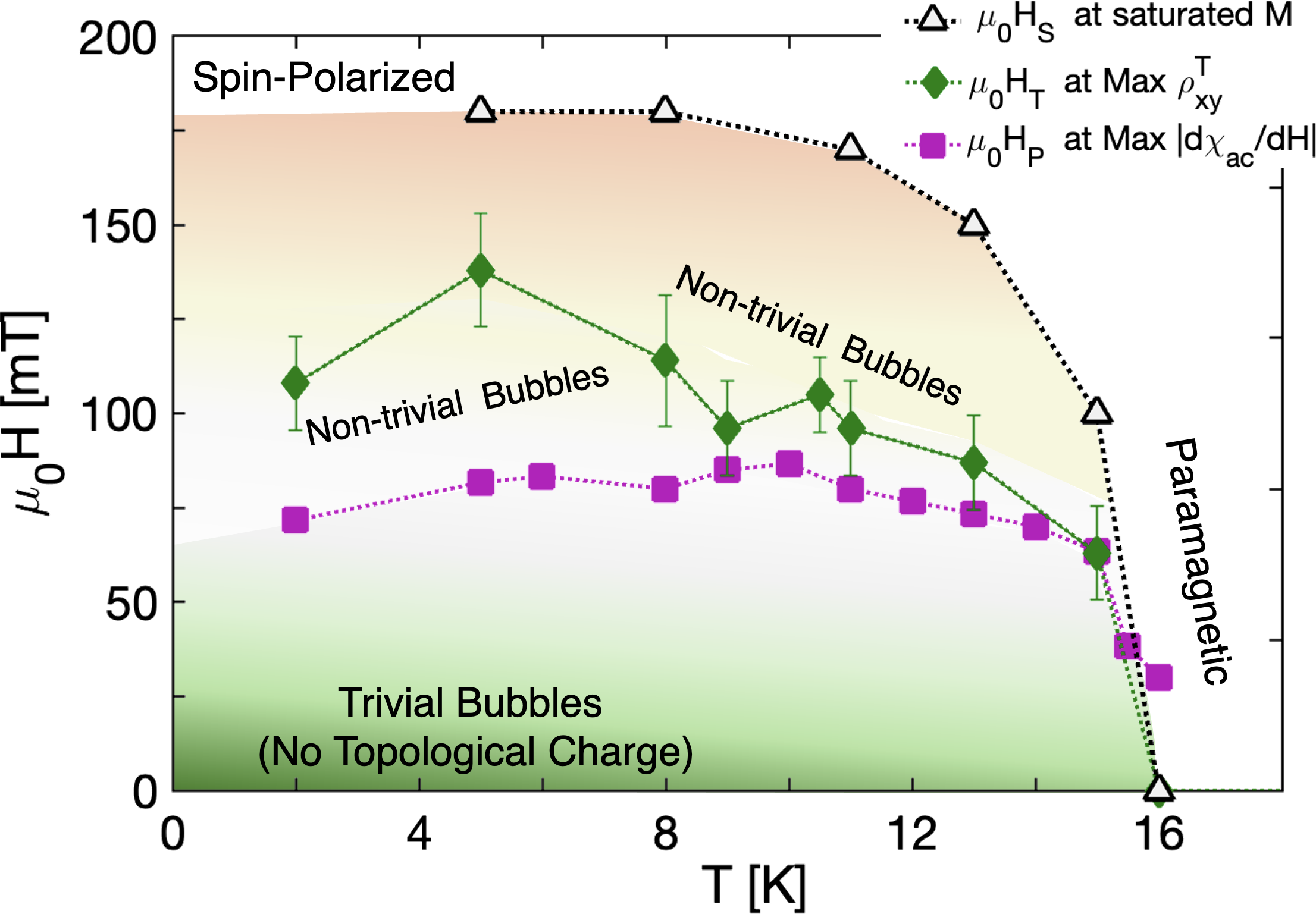}
\caption{\small  
Schematic $H$-$T$  phase diagram of \crgb~ based on the 
 three characteristic field scales  observed in $T< T_C$ is shown: $H_P$ (square) , $H_T$ (diamond) and $H_S$ refer to the field value at the maximum of  $ |\dchidh (H)| $ [Fig.~\ref{chiac}(b)],  at the maximum magnitude of $\rhoxy^T(H)$ [Fig.~\ref{SHmaxTHE}], and at the saturated magnetization [Fig.~\ref{MagMR}(a)],respectively.
The non-topological bubbles state emerges approximately in $H_P<H<H_S$, centered around $H_T$, though its boundaries are not precisely determined, other than the vicinity of $H_T$. 
\black
}
\label{phasedia}
\end{center}
\end{figure}

The broadened features in $\chiac(H)$ of \crgb  are attributed to two differences from  the  B20 materials: 
First, unlike a skyrmion lattice, the magnetic domains in \crgb~do not have a long-range order.  Such a lack of long-range structure in magnetic bubble phases has been reported in other magnetic systems  \cite{Nakamura2018, Maccariello2018,Vistoli2019}. 
Second, as mentioned above, we speculate that the changes in the spin texture of individual bubbles in \crgb occur over a range of fields, which  inevitably causes the finite widths  in step-function like changes in $\chiac(B)$. 
 We will discuss this more with the MFM data in Fig. \ref{mfmdata}. 
 

Fig.~\ref{chiac}(c) shows the $T$ dependence of $\chiac$ taken at constant fields as shown.
The diverging magnitude of $\chiac$ as $T$ approaches $T_C$
 marks the long-range magnetic ordering, consistent with the data shown in Fig.~\ref{chiac}(a). 
We note that the peak of $\chiac(T)$ near $T_C$ begins to split at an applied field of $\mu_0H = 40$ mT, and this becomes more obvious at $\muh=50$ mT and above.  Increasing $H$, the lower  peak moves to the lower $T$, and the higher one to the higher $T$, making the  split more obvious: 
For example,  at  50 mT, the two peaks are located at $T=14.9$  and 16.1 K, respectively, and  they moved  to 14.3 K and 16.5 K at 120 mT, eventually merging into one peak at 17.4 K at 200 mT.  The heights of peaks are in generial reduced gradually as increasing the field but the lower $T$'s peaks decreases faster, while the ones at higher $T$ increase eventually  become the only peak.

At this point,  we do not have a clear explanation for this splitting,  although we speculate that the emergence of two peaks at a finite field originates from robust pinning with an energy scale comparable to  $\approx k_BT_C$ and thus is likely to survive at elevated temperature, of which the similar behaviors were observed in  B20 magnets \cite{Wilhelm2011,Bannenberg2018}. 
\black

We now turn to the MFM images to compare the area fraction of magnetic bubbles \cite{afnote} and the magnitudes of the THE signals.
In Figs.~\ref{mfmdata}(a)-(d), we display the frequency shifts obtained in the same scanned area of $14\times14$ $\mu$m$^2$  at $\muh=40$ and 100 mT, which correspond to below $H_P$ and above $H_T$, respectively.  
The  MFM images are taken after initial field-cooling through $T_C$ at $\muh = 10$ mT. 
 Panels Fig.~\ref{mfmdata}(e) and (f) show the $H$ dependence of the magnitude of $\rhoxy^T$ (left $y$-axes) and the area fraction (right $y$-axes)  for $T=$ 4.2 and 13 K, respectively. 
At $T=4.2$ K, the bubble domains in circular shapes with diameters ranging 
from 0.4 to 1$\mu$m are distributed in the field of view, as shown in Fig.~\ref{mfmdata}(a).
The  shape of the bubbles remains mostly circular until $H$ reaches 100 mT at 4.2 K. However, at 13 K, the shape evolves from circular at low field [panel (c)] to rod-shaped at 100 m T [panel (d)].  The rod-shaped domains that appear only in the higher field, as seen in  Fig.\ref{mfmdata}(d), act as precursors for the stripy domain walls that appear in the spin-polarized state in $H>H_S$ \cite{Wulferding2017} . Hence,  they are more likely to appear at higher $T$s at a given field due to the reduced spin stiffness. 

We define the area fraction as the ratio of the area occupied by bubbles over the scanned area.  We plot the field dependence of the area fractions in comparison to that of $\rhoxy^T$ in Fig. ~\ref{mfmdata}(e) and (f). 
At 4.2 K, it remains roughly constant at the value of  $0.5\pm0.1$until $H$ reaches $H_T$.  Upon increasing $H$ further, the area fraction begins to collapse between $H_P$ and $H_T$ [Fig.~\ref{mfmdata}(e)].  At 13 K, the area fraction exhibits a similar value of 4.2 K  
with a small slope with field up to $\muh =100$ mT. As  $H$  increases further above $H_T$, shares a similar $H$ dependence of $\rhoxy^T$  of a rapid decline with field. 
The coincidence of the decline of the THE signal and the area fraction suggests that the bubbles with trivial (either collinear or coplanar or both) at low field have evolved to have non-zero topological charge that supports the finite THE signal. 
The limitations  of MFM images --  relatively large lateral resolution and  the restricted sensitivity only to the out-of-plane spin component \cite{Shinjo2000, Wachowiak2002} 
-- prevents us from directly tracking such evolutions and hence results in little variation of the area fraction at low field.  However,  with the comparison of the THE,  
we infer a transformation of the internal spin texture of the magnetic bubble.. 
The correlation between the bubble area fraction and the THE has been explored in other uniaxial magnetic thin films such as  (Ca,Ce)MnO$_3$~\cite{Vistoli2019} and La$_{0.7}$Sr$_{0.3}$Mn$_{1-y}$Ru$_y$O$_3$~\cite{Nakamura2018}. 

This observation concurs with the hypothesis that the THE of \crgb~ is enabled by the magnetic bubbles with non-trivial texture present that is gradually developed with the field.  In other words, their total topological charges increase from zero at ZF  and hit a maximum in the narrow range between $H_P$ and $H_T$  and then go back to zero as spin polarization sets in. 
The peak width of $\dchidh$  at $H_P$ marks the wide distribution of the field scales for individual bubbles,   at which the finite topological charge takes shape.


 The next question is what the possible spin configurations would look like
  inside the trivial bubbles that do not contribute to the THE.  The simplest arrangement would be either co-planar or collinear.  
We speculate that,  with the uniaxial anisotropy,  there are two plausible geometries for such spin arrangements: The most straightforward one would be 180$\upcirc$ (Ising-like) domain walls. Here, the characteristic wall width is proportional to $\sqrt{A/K}$, with $A \simeq 2\pi M_S^2$ referring to the ferromagnetic exchange stiffness, and $K$ being the magneto-crystalline anisotropy constant \cite{Niitsu2018}. The condition of $ \mu_0M_S^2/K \ll 1$ for the Ising-like magnet \crgb~ implies a relatively small domain width \cite{Nakamura2018, Vistoli2019}, where the non-trivial spin texture may emerge along with the finite THE signal.
 
 The other spin configuration consists of coplanar magnetic vortex cores \cite{Shinjo2000,Raabe2000} with swirling patterns, \ie non-collinear spin arrangement \cite{Shinjo2000,Raabe2000,Wachowiak2002,Mesler2012}.  
This type is referred to as a magnetic vortex with negligible out-of-plane components, which makes it invisible to the MFM experiments. 
Upon increasing the field, they acquire the out-of-plane components gradually as seen in  Fig. \ref{MagMR} (a), aided by the uniaxial anisotropy, which likely evolves gradually into non-collinear and non-coplanar textures to produce a finite signal.

Finally, we display the schematic phase diagram for the bubble domains as a function of $H$ and $T$ in  Fig. \ref{phasedia},  based on the field dependence of the area fractions and THE in \crgb.
At zero field, bubbles are topologically trivial, but small fields induce nonzero topological charge with widely distributed onset scales. This is reflected in the gradual rise of THE and the broad width of $\dchidh$, of which maximum fall into the close proximity to $ H_T$. 
Near $H_T$, most bubbles transform to having topological texture, which MFM images do not have sensitivity to pick out this change, and thus the area fraction remains constant from zero field to approximately $H_T$.  In  $H>H_T$,  the spin polarization starts and both area fraction and $\rho_{xy}^T$ diminish simultaneously.

To gain insight, we compare  Fig. \ref{phasedia} to the phase diagrams of the B20-structured magnets hosting  the skymion lattice \cite{Bauer2012,Bannenberg2018,Wilhelm2011}.
The onset of the skyrmion lattice phase boundary marked by the THE and 
$\chiac$ in B20 systems is rather abrupt as a function of field, and it is independent of $T$ \cite{Bauer2012}. 
Similarly, in \crgb,  the representative field scales $H_T$ and $H_P$, indicating the onset of the topological bubble domains, remain also mostly $T$-independent until very close to $T_C$, as shown in Fig.~\ref{phasedia}. 
\black

However, the stark difference in \crgb~ is that the onset field to establish topological charge hugely varies bubble by bubble, thus the field dependence of the THE exhibit the gradual rise and fall.  The size of the THE is also determined by the skyrmion density, which is proportional to the effective gauge field responsible for generating the THE \cite{Chapman2013}.  For \crgb, the large bubble size effectively reduces the magnitude of the gauge field, leading to the small magnitude of the  THE signal as well.

\section{Summary} 
We examine the field and temperature dependence of the THE signals and $\chiac$ in 
the centrosymmetric  \crgb~with  an Ising-like uniaxial anisotropy. 
We detect finite THE signals in $2 <T<T_C = 15.5$ K. 
At a fixed $T$, starting from zero field, $\rhoxyT$ increases smoothly from zero upon increasing the field,  having a maximum at $H_T$ before decreasing to zero as  $H$ approaches the saturation field.  
 $\chiac (H) $  exhibits a smooth monotonic field dependence  with $|\dchidh|$  displaying a broad peak with  a maximum at $H_P$, which lies within  close vicinity to $H_T$. 
By analogy with the behavior of $\chi_{ac}(H)$ and $\partial\chi/\partial H$ in B20 magnets, 
we interpret this as a gradual evolution of the spin texture within a magnetic bubble: topologically trivial magnetic bubbles at zero and low fields, which, with increasing field,  transform into topological bubbles carrying nonzero topological charge and thereby contributing to the THE.
Because of the limited sensitivity of magnetic force microscopy to in-plane moments,  it may not capture this evolution, resulting in a constant bubble area fraction up to fields near $H_P$ and $H_T$.
However,  for $H>H_T$, we observe that the reduction of the bubble area fraction, caused by the spin-polarization, follows the same field dependence as the THE. This correspondence indicates that most bubbles acquire topological charge near $H_P$ and $H_T$, although the precise field at which this transformation occurs varies significantly from bubble to bubble, reflected in the broad peak in  $\dchidh$.

While a direct experimental probe of the spin configuration on a sub-micrometer length is not readily accessible for bulk crystals, our result establishes experimentally tangible signatures for the emergence of topological bubbles in the centrosymmetric, DMI-free magnets. 
It is expected to be widely applicable to searches for optimized materials that offer platforms for novel skyrmion-based electronic devices.

\vspace{0.25in}

\begin{acknowledgements}
The work done in the University of Colorado Boulder was supported by the U.S. DOE, Basic Energy Sciences, Materials Sciences and Engineering Division under Award No. DE-SC0021377. 
J. K. was supported by the Ministry of Education, Science, and Technology, South Korea, under Nos. NRF-2016K1A4A01922028, NRF-2018R1A5A6075964, and NRF-2019R1A2C2090356.
D.W. was supported by the faculty research fund of Sejong University in 2025.
E.D.B. was supported by the U.S. DOE, Basic Energy Sciences, Materials Sciences and Engineering Division under the project `quantum fluctuations in narrow band systems'.
F.R. was supported by the U.S. Department of Energy, Office of Science, National Quantum Information Science Research Centers, Quantum Science Center.
\end{acknowledgements}

\vspace{0.25in}

Current Address: 

\noindent $^{\dag}$ Quantinuum,  Broomfield, CO 80021

\noindent
$^{\ddag}$ Department of Physics, University of California, Berkeley, CA 94720

\begin{appendix}
\vspace{0.1 in}

\begin{center}
{\large \bf Appendix }
\end{center}

\begin{figure}
\includegraphics[width=0.9\linewidth]{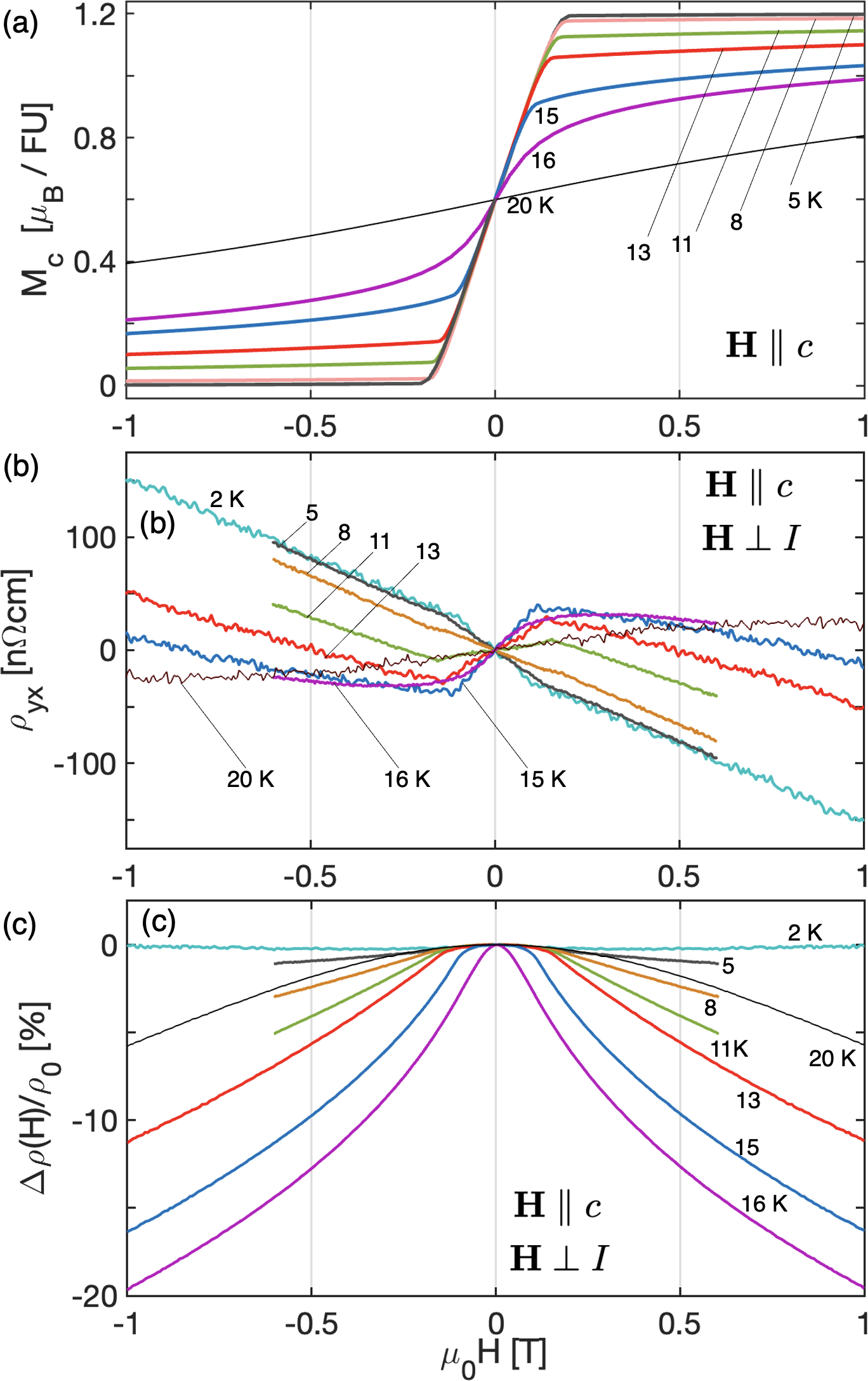}
\caption{
(a) Magnetization ($M_c$)  with the field applied along the $c$-axis ($\mathbf H \parallel c$) up to 1 T at various temperatures as shown. The crystalline $c$-axis is the magnetic easy axis, and the saturation of magnetization is clearly visible at a lower temperature.  
(b) The  Hall resistivity as a function of extended field up to 1T along the $c$-axis  at the $T$'s  shown. The current lies on the $ab$-plane, perpendicular to the applied field.  
(c) Fractional magnetoresistance up to 1 T with 
\black
}
\label{fig:A1}
\end{figure}

In Fig. \ref{fig:A1}, we display  the field dependence of the magnetization, the Hall resisvity and the fMR up to $\pm$ 1T at $T$'s indicated. All panels show the field applied to the $c$-axis (easy axis). 

\end{appendix}
\black

\clearpage


\begin{thebibliography}{73}%
\makeatletter
\providecommand \@ifxundefined [1]{%
 \@ifx{#1\undefined}
}%
\providecommand \@ifnum [1]{%
 \ifnum #1\expandafter \@firstoftwo
 \else \expandafter \@secondoftwo
 \fi
}%
\providecommand \@ifx [1]{%
 \ifx #1\expandafter \@firstoftwo
 \else \expandafter \@secondoftwo
 \fi
}%
\providecommand \natexlab [1]{#1}%
\providecommand \enquote  [1]{``#1''}%
\providecommand \bibnamefont  [1]{#1}%
\providecommand \bibfnamefont [1]{#1}%
\providecommand \citenamefont [1]{#1}%
\providecommand \href@noop [0]{\@secondoftwo}%
\providecommand \href [0]{\begingroup \@sanitize@url \@href}%
\providecommand \@href[1]{\@@startlink{#1}\@@href}%
\providecommand \@@href[1]{\endgroup#1\@@endlink}%
\providecommand \@sanitize@url [0]{\catcode `\\12\catcode `\$12\catcode
  `\&12\catcode `\#12\catcode `\^12\catcode `\_12\catcode `\%12\relax}%
\providecommand \@@startlink[1]{}%
\providecommand \@@endlink[0]{}%
\providecommand \url  [0]{\begingroup\@sanitize@url \@url }%
\providecommand \@url [1]{\endgroup\@href {#1}{\urlprefix }}%
\providecommand \urlprefix  [0]{URL }%
\providecommand \Eprint [0]{\href }%
\providecommand \doibase [0]{http://dx.doi.org/}%
\providecommand \selectlanguage [0]{\@gobble}%
\providecommand \bibinfo  [0]{\@secondoftwo}%
\providecommand \bibfield  [0]{\@secondoftwo}%
\providecommand \translation [1]{[#1]}%
\providecommand \BibitemOpen [0]{}%
\providecommand \bibitemStop [0]{}%
\providecommand \bibitemNoStop [0]{.\EOS\space}%
\providecommand \EOS [0]{\spacefactor3000\relax}%
\providecommand \BibitemShut  [1]{\csname bibitem#1\endcsname}%
\let\auto@bib@innerbib\@empty
\bibitem [{\citenamefont {Fushiya}\ \emph {et~al.}(2014)\citenamefont
  {Fushiya}, \citenamefont {Matsuda}, \citenamefont {Higashinaka},
  \citenamefont {Akiyama},\ and\ \citenamefont {Aoki}}]{Fushiya2014}%
  \BibitemOpen
  \bibfield  {author} {\bibinfo {author} {\bibfnamefont {Kengo}\ \bibnamefont
  {Fushiya}}, \bibinfo {author} {\bibfnamefont {Tatsuma~D.}\ \bibnamefont
  {Matsuda}}, \bibinfo {author} {\bibfnamefont {Ryuji}\ \bibnamefont
  {Higashinaka}}, \bibinfo {author} {\bibfnamefont {Kazuhiro}\ \bibnamefont
  {Akiyama}}, \ and\ \bibinfo {author} {\bibfnamefont {Yuji}\ \bibnamefont
  {Aoki}},\ }\bibfield  {title} {\enquote {\bibinfo {title} {Possible existence
  of partially disordered sm ions in magnetically ordered state of ising magnet
  smpt2si2: A single crystal study},}\ }\href {\doibase 10.7566/JPSJ.83.113708}
  {\bibfield  {journal} {\bibinfo  {journal} {Journal of the Physical Society
  of Japan}\ }\textbf {\bibinfo {volume} {83}},\ \bibinfo {pages} {113708}
  (\bibinfo {year} {2014})}\BibitemShut {NoStop}%
\bibitem [{\citenamefont {Sunku}\ \emph {et~al.}(2016)\citenamefont {Sunku},
  \citenamefont {Kong}, \citenamefont {Ito}, \citenamefont {Canfield},
  \citenamefont {Shastry}, \citenamefont {Sengupta},\ and\ \citenamefont
  {Panagopoulos}}]{Sunku2016}%
  \BibitemOpen
  \bibfield  {author} {\bibinfo {author} {\bibfnamefont {Sai~Swaroop}\
  \bibnamefont {Sunku}}, \bibinfo {author} {\bibfnamefont {Tai}\ \bibnamefont
  {Kong}}, \bibinfo {author} {\bibfnamefont {Toshimitsu}\ \bibnamefont {Ito}},
  \bibinfo {author} {\bibfnamefont {Paul~C.}\ \bibnamefont {Canfield}},
  \bibinfo {author} {\bibfnamefont {B.~Sriram}\ \bibnamefont {Shastry}},
  \bibinfo {author} {\bibfnamefont {Pinaki}\ \bibnamefont {Sengupta}}, \ and\
  \bibinfo {author} {\bibfnamefont {Christos}\ \bibnamefont {Panagopoulos}},\
  }\bibfield  {title} {\enquote {\bibinfo {title} {Hysteretic magnetoresistance
  and unconventional anomalous hall effect in the frustrated magnet
  ${\mathrm{tmb}}_{4}$},}\ }\href {\doibase 10.1103/PhysRevB.93.174408}
  {\bibfield  {journal} {\bibinfo  {journal} {Phys. Rev. B}\ }\textbf {\bibinfo
  {volume} {93}},\ \bibinfo {pages} {174408} (\bibinfo {year}
  {2016})}\BibitemShut {NoStop}%
\bibitem [{\citenamefont {Ye}\ \emph {et~al.}(2017)\citenamefont {Ye},
  \citenamefont {Suzuki},\ and\ \citenamefont {Checkelsky}}]{LYe2017}%
  \BibitemOpen
  \bibfield  {author} {\bibinfo {author} {\bibfnamefont {Linda}\ \bibnamefont
  {Ye}}, \bibinfo {author} {\bibfnamefont {Takehito}\ \bibnamefont {Suzuki}}, \
  and\ \bibinfo {author} {\bibfnamefont {Joseph~G.}\ \bibnamefont
  {Checkelsky}},\ }\bibfield  {title} {\enquote {\bibinfo {title} {Electronic
  transport on the shastry-sutherland lattice in ising-type rare-earth
  tetraborides},}\ }\href {\doibase 10.1103/PhysRevB.95.174405} {\bibfield
  {journal} {\bibinfo  {journal} {Phys. Rev. B}\ }\textbf {\bibinfo {volume}
  {95}},\ \bibinfo {pages} {174405} (\bibinfo {year} {2017})}\BibitemShut
  {NoStop}%
\bibitem [{\citenamefont {L\"ohneysen}\ \emph {et~al.}(2007)\citenamefont
  {L\"ohneysen}, \citenamefont {Rosch}, \citenamefont {Vojta},\ and\
  \citenamefont {W\"olfle}}]{Lohneysen2007_RMP}%
  \BibitemOpen
  \bibfield  {author} {\bibinfo {author} {\bibfnamefont {Hilbert~v.}\
  \bibnamefont {L\"ohneysen}}, \bibinfo {author} {\bibfnamefont {Achim}\
  \bibnamefont {Rosch}}, \bibinfo {author} {\bibfnamefont {Matthias}\
  \bibnamefont {Vojta}}, \ and\ \bibinfo {author} {\bibfnamefont {Peter}\
  \bibnamefont {W\"olfle}},\ }\bibfield  {title} {\enquote {\bibinfo {title}
  {Fermi-liquid instabilities at magnetic quantum phase transitions},}\ }\href
  {\doibase 10.1103/RevModPhys.79.1015} {\bibfield  {journal} {\bibinfo
  {journal} {Rev. Mod. Phys.}\ }\textbf {\bibinfo {volume} {79}},\ \bibinfo
  {pages} {1015--1075} (\bibinfo {year} {2007})}\BibitemShut {NoStop}%
\bibitem [{\citenamefont {Mydosh}\ and\ \citenamefont
  {Oppeneer}(2011)}]{Mydosh2011}%
  \BibitemOpen
  \bibfield  {author} {\bibinfo {author} {\bibfnamefont {J.~A.}\ \bibnamefont
  {Mydosh}}\ and\ \bibinfo {author} {\bibfnamefont {P.~M.}\ \bibnamefont
  {Oppeneer}},\ }\bibfield  {title} {\enquote {\bibinfo {title} {Colloquium:
  Hidden order, superconductivity, and magnetism: The unsolved case of
  ${\mathrm{uru}}_{2}{\mathrm{si}}_{2}$},}\ }\href {\doibase
  10.1103/RevModPhys.83.1301} {\bibfield  {journal} {\bibinfo  {journal} {Rev.
  Mod. Phys.}\ }\textbf {\bibinfo {volume} {83}},\ \bibinfo {pages}
  {1301--1322} (\bibinfo {year} {2011})}\BibitemShut {NoStop}%
\bibitem [{\citenamefont {Pfleiderer}(2009)}]{Pfleiderer2009_RMP}%
  \BibitemOpen
  \bibfield  {author} {\bibinfo {author} {\bibfnamefont {Christian}\
  \bibnamefont {Pfleiderer}},\ }\bibfield  {title} {\enquote {\bibinfo {title}
  {Superconducting phases of $f$-electron compounds},}\ }\href {\doibase
  10.1103/RevModPhys.81.1551} {\bibfield  {journal} {\bibinfo  {journal} {Rev.
  Mod. Phys.}\ }\textbf {\bibinfo {volume} {81}},\ \bibinfo {pages}
  {1551--1624} (\bibinfo {year} {2009})}\BibitemShut {NoStop}%
\bibitem [{\citenamefont {Ruderman}\ and\ \citenamefont
  {Kittel}(1954)}]{Ruderman1954}%
  \BibitemOpen
  \bibfield  {author} {\bibinfo {author} {\bibfnamefont {M.~A.}\ \bibnamefont
  {Ruderman}}\ and\ \bibinfo {author} {\bibfnamefont {C.}~\bibnamefont
  {Kittel}},\ }\bibfield  {title} {\enquote {\bibinfo {title} {Indirect
  exchange coupling of nuclear magnetic moments by conduction electrons},}\
  }\href {\doibase 10.1103/PhysRev.96.99} {\bibfield  {journal} {\bibinfo
  {journal} {Phys. Rev.}\ }\textbf {\bibinfo {volume} {96}},\ \bibinfo {pages}
  {99--102} (\bibinfo {year} {1954})}\BibitemShut {NoStop}%
\bibitem [{\citenamefont {Kasuya}(1956)}]{Kasuya1956}%
  \BibitemOpen
  \bibfield  {author} {\bibinfo {author} {\bibfnamefont {Tadao}\ \bibnamefont
  {Kasuya}},\ }\bibfield  {title} {\enquote {\bibinfo {title} {A theory of
  metallic ferro- and antiferromagnetism on zener's model},}\ }\href {\doibase
  10.1143/PTP.16.45} {\bibfield  {journal} {\bibinfo  {journal} {Progress of
  Theoretical Physics}\ }\textbf {\bibinfo {volume} {16}},\ \bibinfo {pages}
  {45--57} (\bibinfo {year} {1956})},\ \Eprint
  {http://arxiv.org/abs/https://academic.oup.com/ptp/article-pdf/16/1/45/5266722/16-1-45.pdf}
  {https://academic.oup.com/ptp/article-pdf/16/1/45/5266722/16-1-45.pdf}
  \BibitemShut {NoStop}%
\bibitem [{\citenamefont {Yosida}(1957{\natexlab{a}})}]{Yoshida1957}%
  \BibitemOpen
  \bibfield  {author} {\bibinfo {author} {\bibfnamefont {Kei}\ \bibnamefont
  {Yosida}},\ }\bibfield  {title} {\enquote {\bibinfo {title} {Magnetic
  properties of cu-mn alloys},}\ }\href {\doibase 10.1103/PhysRev.106.893}
  {\bibfield  {journal} {\bibinfo  {journal} {Phys. Rev.}\ }\textbf {\bibinfo
  {volume} {106}},\ \bibinfo {pages} {893--898} (\bibinfo {year}
  {1957}{\natexlab{a}})}\BibitemShut {NoStop}%
\bibitem [{\citenamefont {Newman}(1971)}]{Newman1971}%
  \BibitemOpen
  \bibfield  {author} {\bibinfo {author} {\bibfnamefont {D.J.}\ \bibnamefont
  {Newman}},\ }\bibfield  {title} {\enquote {\bibinfo {title} {Theory of
  lanthanide crystal fields},}\ }\href {\doibase 10.1080/00018737100101241}
  {\bibfield  {journal} {\bibinfo  {journal} {Advances in Physics}\ }\textbf
  {\bibinfo {volume} {20}},\ \bibinfo {pages} {197--256} (\bibinfo {year}
  {1971})}\BibitemShut {NoStop}%
\bibitem [{\citenamefont {Shin}\ \emph {et~al.}(2020)\citenamefont {Shin},
  \citenamefont {Pomjakushin}, \citenamefont {Keller}, \citenamefont {Rosa},
  \citenamefont {Stuhr}, \citenamefont {Niedermayer}, \citenamefont {Sibille},
  \citenamefont {Toth}, \citenamefont {Kim}, \citenamefont {Jang},
  \citenamefont {Son}, \citenamefont {Lee}, \citenamefont {Shang},
  \citenamefont {Medarde}, \citenamefont {Bauer}, \citenamefont {Kenzelmann},\
  and\ \citenamefont {Park}}]{Shin2020}%
  \BibitemOpen
  \bibfield  {author} {\bibinfo {author} {\bibfnamefont {S.}~\bibnamefont
  {Shin}}, \bibinfo {author} {\bibfnamefont {V.}~\bibnamefont {Pomjakushin}},
  \bibinfo {author} {\bibfnamefont {L.}~\bibnamefont {Keller}}, \bibinfo
  {author} {\bibfnamefont {P.~F.~S.}\ \bibnamefont {Rosa}}, \bibinfo {author}
  {\bibfnamefont {U.}~\bibnamefont {Stuhr}}, \bibinfo {author} {\bibfnamefont
  {C.}~\bibnamefont {Niedermayer}}, \bibinfo {author} {\bibfnamefont
  {R.}~\bibnamefont {Sibille}}, \bibinfo {author} {\bibfnamefont
  {S.}~\bibnamefont {Toth}}, \bibinfo {author} {\bibfnamefont {J.}~\bibnamefont
  {Kim}}, \bibinfo {author} {\bibfnamefont {H.}~\bibnamefont {Jang}}, \bibinfo
  {author} {\bibfnamefont {S.-K.}\ \bibnamefont {Son}}, \bibinfo {author}
  {\bibfnamefont {H.-O.}\ \bibnamefont {Lee}}, \bibinfo {author} {\bibfnamefont
  {T.}~\bibnamefont {Shang}}, \bibinfo {author} {\bibfnamefont
  {M.}~\bibnamefont {Medarde}}, \bibinfo {author} {\bibfnamefont {E.~D.}\
  \bibnamefont {Bauer}}, \bibinfo {author} {\bibfnamefont {M.}~\bibnamefont
  {Kenzelmann}}, \ and\ \bibinfo {author} {\bibfnamefont {T.}~\bibnamefont
  {Park}},\ }\bibfield  {title} {\enquote {\bibinfo {title} {Magnetic structure
  and crystalline electric field effects in the triangular antiferromagnet
  $\mathrm{CePtAl}_{4}\mathrm{Ge}_{2}$},}\ }\href {\doibase
  10.1103/PhysRevB.101.224421} {\bibfield  {journal} {\bibinfo  {journal}
  {Phys. Rev. B}\ }\textbf {\bibinfo {volume} {101}},\ \bibinfo {pages}
  {224421} (\bibinfo {year} {2020})}\BibitemShut {NoStop}%
\bibitem [{\citenamefont {Kondo}(1964)}]{Kondo1964}%
  \BibitemOpen
  \bibfield  {author} {\bibinfo {author} {\bibfnamefont {Jun}\ \bibnamefont
  {Kondo}},\ }\bibfield  {title} {\enquote {\bibinfo {title} {Resistance
  minimum in dilute magnetic alloys},}\ }\href {\doibase
  doi.org/10.1038/s41467-018-05801-5} {\bibfield  {journal} {\bibinfo
  {journal} {Progress of Theoretical Physics}\ }\textbf {\bibinfo {volume}
  {32}},\ \bibinfo {pages} {37} (\bibinfo {year} {1964})}\BibitemShut {NoStop}%
\bibitem [{\citenamefont {Terashima}\ \emph {et~al.}(2002)\citenamefont
  {Terashima}, \citenamefont {Terakura}, \citenamefont {Uji}, \citenamefont
  {Aoki}, \citenamefont {Echizen},\ and\ \citenamefont
  {Takabatake}}]{Terashima2002}%
  \BibitemOpen
  \bibfield  {author} {\bibinfo {author} {\bibfnamefont {T.}~\bibnamefont
  {Terashima}}, \bibinfo {author} {\bibfnamefont {C.}~\bibnamefont {Terakura}},
  \bibinfo {author} {\bibfnamefont {S.}~\bibnamefont {Uji}}, \bibinfo {author}
  {\bibfnamefont {H.}~\bibnamefont {Aoki}}, \bibinfo {author} {\bibfnamefont
  {Y.}~\bibnamefont {Echizen}}, \ and\ \bibinfo {author} {\bibfnamefont
  {T.}~\bibnamefont {Takabatake}},\ }\bibfield  {title} {\enquote {\bibinfo
  {title} {Resistivity, hall effect, and shubnikov--de haas oscillations in
  cenisn},}\ }\href {\doibase 10.1103/PhysRevB.66.075127} {\bibfield  {journal}
  {\bibinfo  {journal} {Phys. Rev. B}\ }\textbf {\bibinfo {volume} {66}},\
  \bibinfo {pages} {075127} (\bibinfo {year} {2002})}\BibitemShut {NoStop}%
\bibitem [{\citenamefont {Seiro}\ \emph {et~al.}(2018)\citenamefont {Seiro},
  \citenamefont {Jiao}, \citenamefont {Kirchner}, \citenamefont {Hartmann},
  \citenamefont {Friedemann}, \citenamefont {Krellner}, \citenamefont {Geibel},
  \citenamefont {Si}, \citenamefont {Steglich},\ and\ \citenamefont
  {Wirth}}]{Seiro2018}%
  \BibitemOpen
  \bibfield  {author} {\bibinfo {author} {\bibfnamefont {S.}~\bibnamefont
  {Seiro}}, \bibinfo {author} {\bibfnamefont {L.}~\bibnamefont {Jiao}},
  \bibinfo {author} {\bibfnamefont {S.}~\bibnamefont {Kirchner}}, \bibinfo
  {author} {\bibfnamefont {S.}~\bibnamefont {Hartmann}}, \bibinfo {author}
  {\bibfnamefont {S.}~\bibnamefont {Friedemann}}, \bibinfo {author}
  {\bibfnamefont {C.}~\bibnamefont {Krellner}}, \bibinfo {author}
  {\bibfnamefont {C.}~\bibnamefont {Geibel}}, \bibinfo {author} {\bibfnamefont
  {Q.}~\bibnamefont {Si}}, \bibinfo {author} {\bibfnamefont {F.}~\bibnamefont
  {Steglich}}, \ and\ \bibinfo {author} {\bibfnamefont {S.}~\bibnamefont
  {Wirth}},\ }\bibfield  {title} {\enquote {\bibinfo {title} {Evolution of the
  kondo lattice and non-fermi liquid excitations in a heavy-fermion metal},}\
  }\href {\doibase doi.org/10.1038/s41467-018-05801-5} {\bibfield  {journal}
  {\bibinfo  {journal} {Nature Materials}\ }\textbf {\bibinfo {volume} {9}},\
  \bibinfo {pages} {3324} (\bibinfo {year} {2018})}\BibitemShut {NoStop}%
\bibitem [{\citenamefont {Lin}\ and\ \citenamefont {Batista}(2018)}]{Lin2018}%
  \BibitemOpen
  \bibfield  {author} {\bibinfo {author} {\bibfnamefont {Shi-Zeng}\
  \bibnamefont {Lin}}\ and\ \bibinfo {author} {\bibfnamefont {Cristian~D.}\
  \bibnamefont {Batista}},\ }\bibfield  {title} {\enquote {\bibinfo {title}
  {Face centered cubic and hexagonal close packed skyrmion crystals in
  centrosymmetric magnets},}\ }\href {\doibase 10.1103/PhysRevLett.120.077202}
  {\bibfield  {journal} {\bibinfo  {journal} {Phys. Rev. Lett.}\ }\textbf
  {\bibinfo {volume} {120}},\ \bibinfo {pages} {077202} (\bibinfo {year}
  {2018})}\BibitemShut {NoStop}%
\bibitem [{\citenamefont {Kurumajii}\ \emph {et~al.}(2019)\citenamefont
  {Kurumajii}, \citenamefont {Nakajima}, \citenamefont {Hirschberger},
  \citenamefont {Kikkawa}, \citenamefont {Yamasaki}, \citenamefont {Sagayama},
  \citenamefont {Nakao}, \citenamefont {hisa Arima},\ and\ \citenamefont
  {Tokura}}]{Kurumajii2019}%
  \BibitemOpen
  \bibfield  {author} {\bibinfo {author} {\bibfnamefont {Takashi}\ \bibnamefont
  {Kurumajii}}, \bibinfo {author} {\bibfnamefont {Taro}\ \bibnamefont
  {Nakajima}}, \bibinfo {author} {\bibfnamefont {Max}\ \bibnamefont
  {Hirschberger}}, \bibinfo {author} {\bibfnamefont {Akiko}\ \bibnamefont
  {Kikkawa}}, \bibinfo {author} {\bibfnamefont {Yuichi}\ \bibnamefont
  {Yamasaki}}, \bibinfo {author} {\bibfnamefont {Hajime}\ \bibnamefont
  {Sagayama}}, \bibinfo {author} {\bibfnamefont {Hironori}\ \bibnamefont
  {Nakao}}, \bibinfo {author} {\bibfnamefont {Yasujiro Taguchi~Taka}\
  \bibnamefont {hisa Arima}}, \ and\ \bibinfo {author} {\bibfnamefont
  {Yoshinori}\ \bibnamefont {Tokura}},\ }\bibfield  {title} {\enquote {\bibinfo
  {title} {Skyrmion lattice with a giant topological hall effect in a
  frustrated triangular-lattice magnet},}\ }\href {\doibase
  10.1126/science.aau0968} {\bibfield  {journal} {\bibinfo  {journal}
  {Science}\ }\textbf {\bibinfo {volume} {365}},\ \bibinfo {pages} {914--918}
  (\bibinfo {year} {2019})}\BibitemShut {NoStop}%
\bibitem [{\citenamefont {Hirschberger}\ \emph {et~al.}(2019)\citenamefont
  {Hirschberger}, \citenamefont {Nakajima}, \citenamefont {Gao}, \citenamefont
  {Peng}, \citenamefont {Kikkawa}, \citenamefont {Kurumaji}, \citenamefont
  {Kriener}, \citenamefont {Yamasaki}, \citenamefont {Sagayama}, \citenamefont
  {Nakao}, \citenamefont {Ohishi}, \citenamefont {Kakurai}, \citenamefont
  {Taguchi}, \citenamefont {Yu}, \citenamefont {Arima},\ and\ \citenamefont
  {Tokura}}]{Hirschberger2019}%
  \BibitemOpen
  \bibfield  {author} {\bibinfo {author} {\bibfnamefont {Max}\ \bibnamefont
  {Hirschberger}}, \bibinfo {author} {\bibfnamefont {Taro}\ \bibnamefont
  {Nakajima}}, \bibinfo {author} {\bibfnamefont {Shang}\ \bibnamefont {Gao}},
  \bibinfo {author} {\bibfnamefont {Licong}\ \bibnamefont {Peng}}, \bibinfo
  {author} {\bibfnamefont {Akiko}\ \bibnamefont {Kikkawa}}, \bibinfo {author}
  {\bibfnamefont {Takashi}\ \bibnamefont {Kurumaji}}, \bibinfo {author}
  {\bibfnamefont {Markus}\ \bibnamefont {Kriener}}, \bibinfo {author}
  {\bibfnamefont {Yuichi}\ \bibnamefont {Yamasaki}}, \bibinfo {author}
  {\bibfnamefont {Hajime}\ \bibnamefont {Sagayama}}, \bibinfo {author}
  {\bibfnamefont {Hironori}\ \bibnamefont {Nakao}}, \bibinfo {author}
  {\bibfnamefont {Kazuki}\ \bibnamefont {Ohishi}}, \bibinfo {author}
  {\bibfnamefont {Kazuhisa}\ \bibnamefont {Kakurai}}, \bibinfo {author}
  {\bibfnamefont {Yasujiro}\ \bibnamefont {Taguchi}}, \bibinfo {author}
  {\bibfnamefont {Xiuzhen}\ \bibnamefont {Yu}}, \bibinfo {author}
  {\bibfnamefont {Taka-hisa}\ \bibnamefont {Arima}}, \ and\ \bibinfo {author}
  {\bibfnamefont {Yoshinori}\ \bibnamefont {Tokura}},\ }\bibfield  {title}
  {\enquote {\bibinfo {title} {Skyrmion phase and competing magnetic orders on
  a breathing kagome lattice},}\ }\href {\doibase 10.1038/s41467-019-13675-4}
  {\bibfield  {journal} {\bibinfo  {journal} {Nature Communications}\ }\textbf
  {\bibinfo {volume} {10}},\ \bibinfo {pages} {5831} (\bibinfo {year}
  {2019})}\BibitemShut {NoStop}%
\bibitem [{\citenamefont {Leahy}\ \emph {et~al.}(2022)\citenamefont {Leahy},
  \citenamefont {Feng}, \citenamefont {Dery}, \citenamefont {Baumbach},\ and\
  \citenamefont {Lee}}]{Leahy2022}%
  \BibitemOpen
  \bibfield  {author} {\bibinfo {author} {\bibfnamefont {Ian~A.}\ \bibnamefont
  {Leahy}}, \bibinfo {author} {\bibfnamefont {Keke}\ \bibnamefont {Feng}},
  \bibinfo {author} {\bibfnamefont {Roei}\ \bibnamefont {Dery}}, \bibinfo
  {author} {\bibfnamefont {Ryan}\ \bibnamefont {Baumbach}}, \ and\ \bibinfo
  {author} {\bibfnamefont {Minhyea}\ \bibnamefont {Lee}},\ }\bibfield  {title}
  {\enquote {\bibinfo {title} {Field-induced magnetic states in the metallic
  rare-earth layered triangular antiferromagnet
  ${\mathrm{tbaual}}_{4}{\mathrm{ge}}_{2}$},}\ }\href {\doibase
  10.1103/PhysRevB.106.094426} {\bibfield  {journal} {\bibinfo  {journal}
  {Phys. Rev. B}\ }\textbf {\bibinfo {volume} {106}},\ \bibinfo {pages}
  {094426} (\bibinfo {year} {2022})}\BibitemShut {NoStop}%
\bibitem [{\citenamefont {Fruhling}\ \emph {et~al.}(2024)\citenamefont
  {Fruhling}, \citenamefont {Streeter}, \citenamefont {Mardanya}, \citenamefont
  {Wang}, \citenamefont {Baral}, \citenamefont {Zaharko}, \citenamefont
  {Mazin}, \citenamefont {Chowdhury}, \citenamefont {Ratcliff},\ and\
  \citenamefont {Tafti}}]{Fruhling2024}%
  \BibitemOpen
  \bibfield  {author} {\bibinfo {author} {\bibfnamefont {Kyle}\ \bibnamefont
  {Fruhling}}, \bibinfo {author} {\bibfnamefont {Alenna}\ \bibnamefont
  {Streeter}}, \bibinfo {author} {\bibfnamefont {Sougata}\ \bibnamefont
  {Mardanya}}, \bibinfo {author} {\bibfnamefont {Xiaoping}\ \bibnamefont
  {Wang}}, \bibinfo {author} {\bibfnamefont {Priya}\ \bibnamefont {Baral}},
  \bibinfo {author} {\bibfnamefont {Oksana}\ \bibnamefont {Zaharko}}, \bibinfo
  {author} {\bibfnamefont {Igor~I.}\ \bibnamefont {Mazin}}, \bibinfo {author}
  {\bibfnamefont {Sugata}\ \bibnamefont {Chowdhury}}, \bibinfo {author}
  {\bibfnamefont {William~D.}\ \bibnamefont {Ratcliff}}, \ and\ \bibinfo
  {author} {\bibfnamefont {Fazel}\ \bibnamefont {Tafti}},\ }\bibfield  {title}
  {\enquote {\bibinfo {title} {Topological hall effect induced by chiral
  fluctuations in ${\mathrm{ermn}}_{6}{\mathrm{sn}}_{6}$},}\ }\href {\doibase
  10.1103/PhysRevMaterials.8.094411} {\bibfield  {journal} {\bibinfo  {journal}
  {Phys. Rev. Mater.}\ }\textbf {\bibinfo {volume} {8}},\ \bibinfo {pages}
  {094411} (\bibinfo {year} {2024})}\BibitemShut {NoStop}%
\bibitem [{\citenamefont {Ezawa}(2010)}]{Ezawa2010}%
  \BibitemOpen
  \bibfield  {author} {\bibinfo {author} {\bibfnamefont {Motohiko}\
  \bibnamefont {Ezawa}},\ }\bibfield  {title} {\enquote {\bibinfo {title}
  {Giant skyrmions stabilized by dipole-dipole interactions in thin
  ferromagnetic films},}\ }\href {\doibase 10.1103/PhysRevLett.105.197202}
  {\bibfield  {journal} {\bibinfo  {journal} {Phys. Rev. Lett.}\ }\textbf
  {\bibinfo {volume} {105}},\ \bibinfo {pages} {197202} (\bibinfo {year}
  {2010})}\BibitemShut {NoStop}%
\bibitem [{\citenamefont {Kiselev}\ \emph {et~al.}(2011)\citenamefont
  {Kiselev}, \citenamefont {Bogdanov}, \citenamefont {Sch\"afer},\ and\
  \citenamefont {R\"o\ss{}ler}}]{Kiselev2011}%
  \BibitemOpen
  \bibfield  {author} {\bibinfo {author} {\bibfnamefont {N.~S.}\ \bibnamefont
  {Kiselev}}, \bibinfo {author} {\bibfnamefont {A.~N.}\ \bibnamefont
  {Bogdanov}}, \bibinfo {author} {\bibfnamefont {R.}~\bibnamefont {Sch\"afer}},
  \ and\ \bibinfo {author} {\bibfnamefont {U.~K.}\ \bibnamefont
  {R\"o\ss{}ler}},\ }\bibfield  {title} {\enquote {\bibinfo {title} {Comment on
  ``giant skyrmions stabilized by dipole-dipole interactions in thin
  ferromagnetic films''},}\ }\href {\doibase 10.1103/PhysRevLett.107.179701}
  {\bibfield  {journal} {\bibinfo  {journal} {Phys. Rev. Lett.}\ }\textbf
  {\bibinfo {volume} {107}},\ \bibinfo {pages} {179701} (\bibinfo {year}
  {2011})}\BibitemShut {NoStop}%
\bibitem [{\citenamefont {M{\"u}hlbauer}\ \emph {et~al.}(2009)\citenamefont
  {M{\"u}hlbauer}, \citenamefont {Binz}, \citenamefont {Jonietz}, \citenamefont
  {Pfleiderer}, \citenamefont {Rosch}, \citenamefont {Neubauer}, \citenamefont
  {Georgii},\ and\ \citenamefont {B{\"o}ni}}]{Muhlbauer2009}%
  \BibitemOpen
  \bibfield  {author} {\bibinfo {author} {\bibfnamefont {Sebastian}\
  \bibnamefont {M{\"u}hlbauer}}, \bibinfo {author} {\bibfnamefont {Benedikt}\
  \bibnamefont {Binz}}, \bibinfo {author} {\bibfnamefont {F}~\bibnamefont
  {Jonietz}}, \bibinfo {author} {\bibfnamefont {Christian}\ \bibnamefont
  {Pfleiderer}}, \bibinfo {author} {\bibfnamefont {Achim}\ \bibnamefont
  {Rosch}}, \bibinfo {author} {\bibfnamefont {Anja}\ \bibnamefont {Neubauer}},
  \bibinfo {author} {\bibfnamefont {Robert}\ \bibnamefont {Georgii}}, \ and\
  \bibinfo {author} {\bibfnamefont {Peter}\ \bibnamefont {B{\"o}ni}},\
  }\bibfield  {title} {\enquote {\bibinfo {title} {Skyrmion lattice in a chiral
  magnet},}\ }\href@noop {} {\bibfield  {journal} {\bibinfo  {journal}
  {Science}\ }\textbf {\bibinfo {volume} {323}},\ \bibinfo {pages} {915--919}
  (\bibinfo {year} {2009})}\BibitemShut {NoStop}%
\bibitem [{\citenamefont {Yu}\ \emph {et~al.}(2010)\citenamefont {Yu},
  \citenamefont {Onose}, \citenamefont {Kanazawa}, \citenamefont {Park},
  \citenamefont {Han}, \citenamefont {Matsui}, \citenamefont {Nagaosa},\ and\
  \citenamefont {Tokura}}]{XZYu2010}%
  \BibitemOpen
  \bibfield  {author} {\bibinfo {author} {\bibfnamefont {X.~Z.}\ \bibnamefont
  {Yu}}, \bibinfo {author} {\bibfnamefont {Y.}~\bibnamefont {Onose}}, \bibinfo
  {author} {\bibfnamefont {N.}~\bibnamefont {Kanazawa}}, \bibinfo {author}
  {\bibfnamefont {J.~H.}\ \bibnamefont {Park}}, \bibinfo {author}
  {\bibfnamefont {J.~H.}\ \bibnamefont {Han}}, \bibinfo {author} {\bibfnamefont
  {Y.}~\bibnamefont {Matsui}}, \bibinfo {author} {\bibfnamefont
  {N.}~\bibnamefont {Nagaosa}}, \ and\ \bibinfo {author} {\bibfnamefont
  {Y.}~\bibnamefont {Tokura}},\ }\bibfield  {title} {\enquote {\bibinfo {title}
  {Real-space observation of a two-dimensional skyrmion crystal},}\ }\href
  {\doibase 10.1038/nature09124} {\bibfield  {journal} {\bibinfo  {journal}
  {Nature}\ }\textbf {\bibinfo {volume} {465}},\ \bibinfo {pages} {901--904}
  (\bibinfo {year} {2010})}\BibitemShut {NoStop}%
\bibitem [{\citenamefont {Yu}\ \emph {et~al.}(2011)\citenamefont {Yu},
  \citenamefont {Kanazawa}, \citenamefont {Onose}, \citenamefont {Kimoto},
  \citenamefont {andS. Ishiwata}, \citenamefont {Matsui},\ and\ \citenamefont
  {Tokura}}]{XZYu2011}%
  \BibitemOpen
  \bibfield  {author} {\bibinfo {author} {\bibfnamefont {X.~Z.}\ \bibnamefont
  {Yu}}, \bibinfo {author} {\bibfnamefont {N.}~\bibnamefont {Kanazawa}},
  \bibinfo {author} {\bibfnamefont {Y.}~\bibnamefont {Onose}}, \bibinfo
  {author} {\bibfnamefont {K.}~\bibnamefont {Kimoto}}, \bibinfo {author}
  {\bibfnamefont {W.~Z.~Zhang}\ \bibnamefont {andS. Ishiwata}}, \bibinfo
  {author} {\bibfnamefont {Y.}~\bibnamefont {Matsui}}, \ and\ \bibinfo {author}
  {\bibfnamefont {Y.}~\bibnamefont {Tokura}},\ }\bibfield  {title} {\enquote
  {\bibinfo {title} {Near room-temperature formation of a skyrmion crystal in
  thin-films of the helimagnet fege},}\ }\href {\doibase 10.1038/nmat2916}
  {\bibfield  {journal} {\bibinfo  {journal} {Nature Materials}\ }\textbf
  {\bibinfo {volume} {10}},\ \bibinfo {pages} {106--109} (\bibinfo {year}
  {2011})}\BibitemShut {NoStop}%
\bibitem [{\citenamefont {Piva}\ \emph {et~al.}(2023)\citenamefont {Piva},
  \citenamefont {Souza}, \citenamefont {Lombardi}, \citenamefont {Pakuszewski},
  \citenamefont {Adriano}, \citenamefont {Pagliuso},\ and\ \citenamefont
  {Nicklas}}]{Piva2023}%
  \BibitemOpen
  \bibfield  {author} {\bibinfo {author} {\bibfnamefont {M.~M.}\ \bibnamefont
  {Piva}}, \bibinfo {author} {\bibfnamefont {J.~C.}\ \bibnamefont {Souza}},
  \bibinfo {author} {\bibfnamefont {G.~A.}\ \bibnamefont {Lombardi}}, \bibinfo
  {author} {\bibfnamefont {K.~R.}\ \bibnamefont {Pakuszewski}}, \bibinfo
  {author} {\bibfnamefont {C.}~\bibnamefont {Adriano}}, \bibinfo {author}
  {\bibfnamefont {P.~G.}\ \bibnamefont {Pagliuso}}, \ and\ \bibinfo {author}
  {\bibfnamefont {M.}~\bibnamefont {Nicklas}},\ }\bibfield  {title} {\enquote
  {\bibinfo {title} {Topological hall effect in cealge},}\ }\href {\doibase
  10.1103/PhysRevMaterials.7.074204} {\bibfield  {journal} {\bibinfo  {journal}
  {Phys. Rev. Mater.}\ }\textbf {\bibinfo {volume} {7}},\ \bibinfo {pages}
  {074204} (\bibinfo {year} {2023})}\BibitemShut {NoStop}%
\bibitem [{\citenamefont {R{\"o}{\ss}ler}\ \emph {et~al.}(2006)\citenamefont
  {R{\"o}{\ss}ler}, \citenamefont {Bogdanov},\ and\ \citenamefont
  {Pfleiderer}}]{Rossler2006}%
  \BibitemOpen
  \bibfield  {author} {\bibinfo {author} {\bibfnamefont {UK}~\bibnamefont
  {R{\"o}{\ss}ler}}, \bibinfo {author} {\bibfnamefont {AN}~\bibnamefont
  {Bogdanov}}, \ and\ \bibinfo {author} {\bibfnamefont {C}~\bibnamefont
  {Pfleiderer}},\ }\bibfield  {title} {\enquote {\bibinfo {title} {Spontaneous
  skyrmion ground states in magnetic metals},}\ }\href@noop {} {\bibfield
  {journal} {\bibinfo  {journal} {Nature}\ }\textbf {\bibinfo {volume} {442}},\
  \bibinfo {pages} {797--801} (\bibinfo {year} {2006})}\BibitemShut {NoStop}%
\bibitem [{\citenamefont {Nagaosa}\ and\ \citenamefont
  {Tokura}(2013)}]{Nagaosa2013}%
  \BibitemOpen
  \bibfield  {author} {\bibinfo {author} {\bibfnamefont {Naoto}\ \bibnamefont
  {Nagaosa}}\ and\ \bibinfo {author} {\bibfnamefont {Yoshinori}\ \bibnamefont
  {Tokura}},\ }\bibfield  {title} {\enquote {\bibinfo {title} {Topological
  properties and dynamics of magnetic skyrmions},}\ }\href@noop {} {\bibfield
  {journal} {\bibinfo  {journal} {Nature nanotechnology}\ }\textbf {\bibinfo
  {volume} {8}},\ \bibinfo {pages} {899--911} (\bibinfo {year}
  {2013})}\BibitemShut {NoStop}%
\bibitem [{\citenamefont {Leeuw}\ \emph {et~al.}(1980)\citenamefont {Leeuw},
  \citenamefont {Doel},\ and\ \citenamefont {Enz}}]{Leeuw1980}%
  \BibitemOpen
  \bibfield  {author} {\bibinfo {author} {\bibfnamefont {F~H~De}\ \bibnamefont
  {Leeuw}}, \bibinfo {author} {\bibfnamefont {R~Van~Den}\ \bibnamefont {Doel}},
  \ and\ \bibinfo {author} {\bibfnamefont {U}~\bibnamefont {Enz}},\ }\bibfield
  {title} {\enquote {\bibinfo {title} {Dynamic properties of magnetic domain
  walls and magnetic bubbles},}\ }\href {\doibase 10.1088/0034-4885/43/6/001}
  {\bibfield  {journal} {\bibinfo  {journal} {Reports on Progress in Physics}\
  }\textbf {\bibinfo {volume} {43}},\ \bibinfo {pages} {689} (\bibinfo {year}
  {1980})}\BibitemShut {NoStop}%
\bibitem [{\citenamefont {Li}\ \emph {et~al.}(2013)\citenamefont {Li},
  \citenamefont {Kanazawa}, \citenamefont {Yu}, \citenamefont {Tsukazaki},
  \citenamefont {Kawasaki}, \citenamefont {Ichikawa}, \citenamefont {Jin},
  \citenamefont {Kagawa},\ and\ \citenamefont {Tokura}}]{YFLi2013}%
  \BibitemOpen
  \bibfield  {author} {\bibinfo {author} {\bibfnamefont {Yufan}\ \bibnamefont
  {Li}}, \bibinfo {author} {\bibfnamefont {N.}~\bibnamefont {Kanazawa}},
  \bibinfo {author} {\bibfnamefont {X.~Z.}\ \bibnamefont {Yu}}, \bibinfo
  {author} {\bibfnamefont {A.}~\bibnamefont {Tsukazaki}}, \bibinfo {author}
  {\bibfnamefont {M.}~\bibnamefont {Kawasaki}}, \bibinfo {author}
  {\bibfnamefont {M.}~\bibnamefont {Ichikawa}}, \bibinfo {author}
  {\bibfnamefont {X.~F.}\ \bibnamefont {Jin}}, \bibinfo {author} {\bibfnamefont
  {F.}~\bibnamefont {Kagawa}}, \ and\ \bibinfo {author} {\bibfnamefont
  {Y.}~\bibnamefont {Tokura}},\ }\bibfield  {title} {\enquote {\bibinfo {title}
  {Robust formation of skyrmions and topological hall effect anomaly in
  epitaxial thin films of mnsi},}\ }\href {\doibase
  10.1103/PhysRevLett.110.117202} {\bibfield  {journal} {\bibinfo  {journal}
  {Phys. Rev. Lett.}\ }\textbf {\bibinfo {volume} {110}},\ \bibinfo {pages}
  {117202} (\bibinfo {year} {2013})}\BibitemShut {NoStop}%
\bibitem [{\citenamefont {Fert}\ \emph {et~al.}(2013)\citenamefont {Fert},
  \citenamefont {Cros},\ and\ \citenamefont {Sampaio}}]{Fert2013}%
  \BibitemOpen
  \bibfield  {author} {\bibinfo {author} {\bibfnamefont {Albert}\ \bibnamefont
  {Fert}}, \bibinfo {author} {\bibfnamefont {Vincent}\ \bibnamefont {Cros}}, \
  and\ \bibinfo {author} {\bibfnamefont {Jo{\~a}o}\ \bibnamefont {Sampaio}},\
  }\bibfield  {title} {\enquote {\bibinfo {title} {Skyrmions on the track},}\
  }\href@noop {} {\bibfield  {journal} {\bibinfo  {journal} {Nature
  nanotechnology}\ }\textbf {\bibinfo {volume} {8}},\ \bibinfo {pages}
  {152--156} (\bibinfo {year} {2013})}\BibitemShut {NoStop}%
\bibitem [{\citenamefont {Baumbach}\ \emph {et~al.}(2012)\citenamefont
  {Baumbach}, \citenamefont {Shang}, \citenamefont {Torrez}, \citenamefont
  {Ronning}, \citenamefont {Thompson},\ and\ \citenamefont
  {Bauer}}]{Baumbach2012JoP}%
  \BibitemOpen
  \bibfield  {author} {\bibinfo {author} {\bibfnamefont {R~E}\ \bibnamefont
  {Baumbach}}, \bibinfo {author} {\bibfnamefont {T}~\bibnamefont {Shang}},
  \bibinfo {author} {\bibfnamefont {M}~\bibnamefont {Torrez}}, \bibinfo
  {author} {\bibfnamefont {F}~\bibnamefont {Ronning}}, \bibinfo {author}
  {\bibfnamefont {J~D}\ \bibnamefont {Thompson}}, \ and\ \bibinfo {author}
  {\bibfnamefont {E~D}\ \bibnamefont {Bauer}},\ }\bibfield  {title} {\enquote
  {\bibinfo {title} {Local moment ferromagnetism in {CeRu}2ga2b},}\ }\href
  {\doibase 10.1088/0953-8984/24/18/185702} {\bibfield  {journal} {\bibinfo
  {journal} {Journal of Physics: Condensed Matter}\ }\textbf {\bibinfo {volume}
  {24}},\ \bibinfo {pages} {185702} (\bibinfo {year} {2012})}\BibitemShut
  {NoStop}%
\bibitem [{\citenamefont {Sakai}\ \emph {et~al.}(2012)\citenamefont {Sakai},
  \citenamefont {Tokunaga}, \citenamefont {Kambe}, \citenamefont {Baumbach},
  \citenamefont {Ronning}, \citenamefont {Bauer},\ and\ \citenamefont
  {Thompson}}]{Sakai2012}%
  \BibitemOpen
  \bibfield  {author} {\bibinfo {author} {\bibfnamefont {H.}~\bibnamefont
  {Sakai}}, \bibinfo {author} {\bibfnamefont {Y.}~\bibnamefont {Tokunaga}},
  \bibinfo {author} {\bibfnamefont {S.}~\bibnamefont {Kambe}}, \bibinfo
  {author} {\bibfnamefont {R.~E.}\ \bibnamefont {Baumbach}}, \bibinfo {author}
  {\bibfnamefont {F.}~\bibnamefont {Ronning}}, \bibinfo {author} {\bibfnamefont
  {E.~D.}\ \bibnamefont {Bauer}}, \ and\ \bibinfo {author} {\bibfnamefont
  {J.~D.}\ \bibnamefont {Thompson}},\ }\bibfield  {title} {\enquote {\bibinfo
  {title} {Nmr study for $4f$-localized ferromagnet ceru${}_{2}$ga${}_{2}$b},}\
  }\href {\doibase 10.1103/PhysRevB.86.094402} {\bibfield  {journal} {\bibinfo
  {journal} {Phys. Rev. B}\ }\textbf {\bibinfo {volume} {86}},\ \bibinfo
  {pages} {094402} (\bibinfo {year} {2012})}\BibitemShut {NoStop}%
\bibitem [{\citenamefont {Matsuoka}\ \emph {et~al.}(2012)\citenamefont
  {Matsuoka}, \citenamefont {Tomiyama}, \citenamefont {Sugawara}, \citenamefont
  {Sakurai},\ and\ \citenamefont {Ohta}}]{Matsuoka2012}%
  \BibitemOpen
  \bibfield  {author} {\bibinfo {author} {\bibfnamefont {Eiichi}\ \bibnamefont
  {Matsuoka}}, \bibinfo {author} {\bibfnamefont {Yo}~\bibnamefont {Tomiyama}},
  \bibinfo {author} {\bibfnamefont {Hitoshi}\ \bibnamefont {Sugawara}},
  \bibinfo {author} {\bibfnamefont {Takahiro}\ \bibnamefont {Sakurai}}, \ and\
  \bibinfo {author} {\bibfnamefont {Hitoshi}\ \bibnamefont {Ohta}},\ }\bibfield
   {title} {\enquote {\bibinfo {title} {Ferromagnetic ground states with high
  transition temperatures in new tetragonal rare-earth compounds ceru2al2b and
  prru2al2b},}\ }\href {\doibase 10.1143/JPSJ.81.043704} {\bibfield  {journal}
  {\bibinfo  {journal} {Journal of the Physical Society of Japan}\ }\textbf
  {\bibinfo {volume} {81}},\ \bibinfo {pages} {043704} (\bibinfo {year}
  {2012})},\ \Eprint
  {http://arxiv.org/abs/https://doi.org/10.1143/JPSJ.81.043704}
  {https://doi.org/10.1143/JPSJ.81.043704} \BibitemShut {NoStop}%
\bibitem [{\citenamefont {Wulferding}\ \emph {et~al.}(2017)\citenamefont
  {Wulferding}, \citenamefont {Kim}, \citenamefont {Yang}, \citenamefont
  {Jeong}, \citenamefont {Barros}, \citenamefont {Kato}, \citenamefont
  {Martin}, \citenamefont {Ayala-Valenzuela}, \citenamefont {Lee},
  \citenamefont {Choi} \emph {et~al.}}]{Wulferding2017}%
  \BibitemOpen
  \bibfield  {author} {\bibinfo {author} {\bibfnamefont {Dirk}\ \bibnamefont
  {Wulferding}}, \bibinfo {author} {\bibfnamefont {Hoon}\ \bibnamefont {Kim}},
  \bibinfo {author} {\bibfnamefont {Ilkyu}\ \bibnamefont {Yang}}, \bibinfo
  {author} {\bibfnamefont {Juyoung}\ \bibnamefont {Jeong}}, \bibinfo {author}
  {\bibfnamefont {K}~\bibnamefont {Barros}}, \bibinfo {author} {\bibfnamefont
  {Y}~\bibnamefont {Kato}}, \bibinfo {author} {\bibfnamefont {I}~\bibnamefont
  {Martin}}, \bibinfo {author} {\bibfnamefont {OE}~\bibnamefont
  {Ayala-Valenzuela}}, \bibinfo {author} {\bibfnamefont {Minkyung}\
  \bibnamefont {Lee}}, \bibinfo {author} {\bibfnamefont {Hee~Cheul}\
  \bibnamefont {Choi}},  \emph {et~al.},\ }\bibfield  {title} {\enquote
  {\bibinfo {title} {Domain engineering of the metastable domains in the
  4f-uniaxial-ferromagnet ceru 2 ga 2 b},}\ }\href@noop {} {\bibfield
  {journal} {\bibinfo  {journal} {Scientific reports}\ }\textbf {\bibinfo
  {volume} {7}},\ \bibinfo {pages} {46296} (\bibinfo {year}
  {2017})}\BibitemShut {NoStop}%
\bibitem [{\citenamefont {Hayami}\ and\ \citenamefont
  {Motome}(2018)}]{Hayami2018}%
  \BibitemOpen
  \bibfield  {author} {\bibinfo {author} {\bibfnamefont {Satoru}\ \bibnamefont
  {Hayami}}\ and\ \bibinfo {author} {\bibfnamefont {Yukitoshi}\ \bibnamefont
  {Motome}},\ }\bibfield  {title} {\enquote {\bibinfo {title} {N\'eel- and
  bloch-type magnetic vortices in rashba metals},}\ }\href {\doibase
  10.1103/PhysRevLett.121.137202} {\bibfield  {journal} {\bibinfo  {journal}
  {Phys. Rev. Lett.}\ }\textbf {\bibinfo {volume} {121}},\ \bibinfo {pages}
  {137202} (\bibinfo {year} {2018})}\BibitemShut {NoStop}%
\bibitem [{\citenamefont {Lee}\ \emph {et~al.}(2009)\citenamefont {Lee},
  \citenamefont {Kang}, \citenamefont {Onose}, \citenamefont {Tokura},\ and\
  \citenamefont {Ong}}]{Lee2009}%
  \BibitemOpen
  \bibfield  {author} {\bibinfo {author} {\bibfnamefont {Minhyea}\ \bibnamefont
  {Lee}}, \bibinfo {author} {\bibfnamefont {W.}~\bibnamefont {Kang}}, \bibinfo
  {author} {\bibfnamefont {Y.}~\bibnamefont {Onose}}, \bibinfo {author}
  {\bibfnamefont {Y.}~\bibnamefont {Tokura}}, \ and\ \bibinfo {author}
  {\bibfnamefont {N.~P.}\ \bibnamefont {Ong}},\ }\bibfield  {title} {\enquote
  {\bibinfo {title} {Unusual hall effect anomaly in mnsi under pressure},}\
  }\href {\doibase 10.1103/PhysRevLett.102.186601} {\bibfield  {journal}
  {\bibinfo  {journal} {Phys. Rev. Lett.}\ }\textbf {\bibinfo {volume} {102}},\
  \bibinfo {pages} {186601} (\bibinfo {year} {2009})}\BibitemShut {NoStop}%
\bibitem [{\citenamefont {Neubauer}\ \emph {et~al.}(2009)\citenamefont
  {Neubauer}, \citenamefont {Pfleiderer}, \citenamefont {Binz}, \citenamefont
  {Rosch}, \citenamefont {Ritz}, \citenamefont {Niklowitz},\ and\ \citenamefont
  {B{\"o}ni}}]{Neubauer2009}%
  \BibitemOpen
  \bibfield  {author} {\bibinfo {author} {\bibfnamefont {A}~\bibnamefont
  {Neubauer}}, \bibinfo {author} {\bibfnamefont {C}~\bibnamefont {Pfleiderer}},
  \bibinfo {author} {\bibfnamefont {B}~\bibnamefont {Binz}}, \bibinfo {author}
  {\bibfnamefont {A}~\bibnamefont {Rosch}}, \bibinfo {author} {\bibfnamefont
  {R}~\bibnamefont {Ritz}}, \bibinfo {author} {\bibfnamefont {PG}~\bibnamefont
  {Niklowitz}}, \ and\ \bibinfo {author} {\bibfnamefont {P}~\bibnamefont
  {B{\"o}ni}},\ }\bibfield  {title} {\enquote {\bibinfo {title} {Topological
  hall effect in the a phase of mnsi},}\ }\href@noop {} {\bibfield  {journal}
  {\bibinfo  {journal} {Physical review letters}\ }\textbf {\bibinfo {volume}
  {102}},\ \bibinfo {pages} {186602} (\bibinfo {year} {2009})}\BibitemShut
  {NoStop}%
\bibitem [{\citenamefont {Schulz}\ \emph {et~al.}(2012)\citenamefont {Schulz},
  \citenamefont {Ritz}, \citenamefont {Bauer}, \citenamefont {Halder},
  \citenamefont {Wagner}, \citenamefont {Franz}, \citenamefont {Pfleiderer},
  \citenamefont {Everschor}, \citenamefont {Garst},\ and\ \citenamefont
  {Rosch}}]{Schulz2012}%
  \BibitemOpen
  \bibfield  {author} {\bibinfo {author} {\bibfnamefont {T}~\bibnamefont
  {Schulz}}, \bibinfo {author} {\bibfnamefont {R}~\bibnamefont {Ritz}},
  \bibinfo {author} {\bibfnamefont {A}~\bibnamefont {Bauer}}, \bibinfo {author}
  {\bibfnamefont {M}~\bibnamefont {Halder}}, \bibinfo {author} {\bibfnamefont
  {M}~\bibnamefont {Wagner}}, \bibinfo {author} {\bibfnamefont {C}~\bibnamefont
  {Franz}}, \bibinfo {author} {\bibfnamefont {C}~\bibnamefont {Pfleiderer}},
  \bibinfo {author} {\bibfnamefont {K}~\bibnamefont {Everschor}}, \bibinfo
  {author} {\bibfnamefont {M}~\bibnamefont {Garst}}, \ and\ \bibinfo {author}
  {\bibfnamefont {A}~\bibnamefont {Rosch}},\ }\bibfield  {title} {\enquote
  {\bibinfo {title} {Emergent electrodynamics of skyrmions in a chiral
  magnet},}\ }\href@noop {} {\bibfield  {journal} {\bibinfo  {journal} {Nature
  Physics}\ }\textbf {\bibinfo {volume} {8}},\ \bibinfo {pages} {301--304}
  (\bibinfo {year} {2012})}\BibitemShut {NoStop}%
\bibitem [{\citenamefont {Bauer}\ and\ \citenamefont
  {Pfleiderer}(2012)}]{Bauer2012}%
  \BibitemOpen
  \bibfield  {author} {\bibinfo {author} {\bibfnamefont {A.}~\bibnamefont
  {Bauer}}\ and\ \bibinfo {author} {\bibfnamefont {C.}~\bibnamefont
  {Pfleiderer}},\ }\bibfield  {title} {\enquote {\bibinfo {title} {Magnetic
  phase diagram of mnsi inferred from magnetization and ac susceptibility},}\
  }\href {\doibase 10.1103/PhysRevB.85.214418} {\bibfield  {journal} {\bibinfo
  {journal} {Phys. Rev. B}\ }\textbf {\bibinfo {volume} {85}},\ \bibinfo
  {pages} {214418} (\bibinfo {year} {2012})}\BibitemShut {NoStop}%
\bibitem [{\citenamefont {Bauer}\ \emph {et~al.}(2017)\citenamefont {Bauer},
  \citenamefont {Chacon}, \citenamefont {Wagner}, \citenamefont {Halder},
  \citenamefont {Georgii}, \citenamefont {Rosch}, \citenamefont {Pfleiderer},\
  and\ \citenamefont {Garst}}]{Bauer2017}%
  \BibitemOpen
  \bibfield  {author} {\bibinfo {author} {\bibfnamefont {A.}~\bibnamefont
  {Bauer}}, \bibinfo {author} {\bibfnamefont {A.}~\bibnamefont {Chacon}},
  \bibinfo {author} {\bibfnamefont {M.}~\bibnamefont {Wagner}}, \bibinfo
  {author} {\bibfnamefont {M.}~\bibnamefont {Halder}}, \bibinfo {author}
  {\bibfnamefont {R.}~\bibnamefont {Georgii}}, \bibinfo {author} {\bibfnamefont
  {A.}~\bibnamefont {Rosch}}, \bibinfo {author} {\bibfnamefont
  {C.}~\bibnamefont {Pfleiderer}}, \ and\ \bibinfo {author} {\bibfnamefont
  {M.}~\bibnamefont {Garst}},\ }\bibfield  {title} {\enquote {\bibinfo {title}
  {Symmetry breaking, slow relaxation dynamics, and topological defects at the
  field-induced helix reorientation in mnsi},}\ }\href {\doibase
  10.1103/PhysRevB.95.024429} {\bibfield  {journal} {\bibinfo  {journal} {Phys.
  Rev. B}\ }\textbf {\bibinfo {volume} {95}},\ \bibinfo {pages} {024429}
  (\bibinfo {year} {2017})}\BibitemShut {NoStop}%
\bibitem [{\citenamefont {Bauer}\ \emph {et~al.}(2013)\citenamefont {Bauer},
  \citenamefont {Garst},\ and\ \citenamefont {Pfleiderer}}]{Bauer2013}%
  \BibitemOpen
  \bibfield  {author} {\bibinfo {author} {\bibfnamefont {A.}~\bibnamefont
  {Bauer}}, \bibinfo {author} {\bibfnamefont {M.}~\bibnamefont {Garst}}, \ and\
  \bibinfo {author} {\bibfnamefont {C.}~\bibnamefont {Pfleiderer}},\ }\bibfield
   {title} {\enquote {\bibinfo {title} {Specific heat of the skyrmion lattice
  phase and field-induced tricritical point in mnsi},}\ }\href {\doibase
  10.1103/PhysRevLett.110.177207} {\bibfield  {journal} {\bibinfo  {journal}
  {Phys. Rev. Lett.}\ }\textbf {\bibinfo {volume} {110}},\ \bibinfo {pages}
  {177207} (\bibinfo {year} {2013})}\BibitemShut {NoStop}%
\bibitem [{\citenamefont {Siegfried}\ \emph {et~al.}(2017)\citenamefont
  {Siegfried}, \citenamefont {Bornstein}, \citenamefont {Treglia},
  \citenamefont {Wolf},\ and\ \citenamefont {Lee}}]{Siegfried2017}%
  \BibitemOpen
  \bibfield  {author} {\bibinfo {author} {\bibfnamefont {Peter~E.}\
  \bibnamefont {Siegfried}}, \bibinfo {author} {\bibfnamefont {Alexander~C.}\
  \bibnamefont {Bornstein}}, \bibinfo {author} {\bibfnamefont {Andrew~C.}\
  \bibnamefont {Treglia}}, \bibinfo {author} {\bibfnamefont {Thomas}\
  \bibnamefont {Wolf}}, \ and\ \bibinfo {author} {\bibfnamefont {Minhyea}\
  \bibnamefont {Lee}},\ }\bibfield  {title} {\enquote {\bibinfo {title}
  {Multiple magnetic states within the $a$ phase determined by
  field-orientation dependence of
  ${\mathbf{mn}}_{0.9}{\mathbf{fe}}_{0.1}\mathbf{Si}$},}\ }\href {\doibase
  10.1103/PhysRevB.96.220410} {\bibfield  {journal} {\bibinfo  {journal} {Phys.
  Rev. B}\ }\textbf {\bibinfo {volume} {96}},\ \bibinfo {pages} {220410}
  (\bibinfo {year} {2017})}\BibitemShut {NoStop}%
\bibitem [{\citenamefont {Lobanova}\ \emph {et~al.}(2016)\citenamefont
  {Lobanova}, \citenamefont {Glushkov}, \citenamefont {Sluchanko},\ and\
  \citenamefont {Demishev}}]{Lobanova2016}%
  \BibitemOpen
  \bibfield  {author} {\bibinfo {author} {\bibfnamefont {II}~\bibnamefont
  {Lobanova}}, \bibinfo {author} {\bibfnamefont {VV}~\bibnamefont {Glushkov}},
  \bibinfo {author} {\bibfnamefont {NE}~\bibnamefont {Sluchanko}}, \ and\
  \bibinfo {author} {\bibfnamefont {SV}~\bibnamefont {Demishev}},\ }\bibfield
  {title} {\enquote {\bibinfo {title} {Macroscopic evidence for abrikosov-type
  magnetic vortexes in mnsi a-phase},}\ }\href@noop {} {\bibfield  {journal}
  {\bibinfo  {journal} {Scientific Reports}\ }\textbf {\bibinfo {volume} {6}},\
  \bibinfo {pages} {22101} (\bibinfo {year} {2016})}\BibitemShut {NoStop}%
\bibitem [{\citenamefont {Wilhelm}\ \emph {et~al.}(2011)\citenamefont
  {Wilhelm}, \citenamefont {Baenitz}, \citenamefont {Schmidt}, \citenamefont
  {R\"o\ss{}ler}, \citenamefont {Leonov},\ and\ \citenamefont
  {Bogdanov}}]{Wilhelm2011}%
  \BibitemOpen
  \bibfield  {author} {\bibinfo {author} {\bibfnamefont {H.}~\bibnamefont
  {Wilhelm}}, \bibinfo {author} {\bibfnamefont {M.}~\bibnamefont {Baenitz}},
  \bibinfo {author} {\bibfnamefont {M.}~\bibnamefont {Schmidt}}, \bibinfo
  {author} {\bibfnamefont {U.~K.}\ \bibnamefont {R\"o\ss{}ler}}, \bibinfo
  {author} {\bibfnamefont {A.~A.}\ \bibnamefont {Leonov}}, \ and\ \bibinfo
  {author} {\bibfnamefont {A.~N.}\ \bibnamefont {Bogdanov}},\ }\bibfield
  {title} {\enquote {\bibinfo {title} {Precursor phenomena at the magnetic
  ordering of the cubic helimagnet fege},}\ }\href {\doibase
  10.1103/PhysRevLett.107.127203} {\bibfield  {journal} {\bibinfo  {journal}
  {Phys. Rev. Lett.}\ }\textbf {\bibinfo {volume} {107}},\ \bibinfo {pages}
  {127203} (\bibinfo {year} {2011})}\BibitemShut {NoStop}%
\bibitem [{\citenamefont {Bannenberg}\ \emph {et~al.}(2018)\citenamefont
  {Bannenberg}, \citenamefont {Weber}, \citenamefont {Lefering}, \citenamefont
  {Wolf},\ and\ \citenamefont {Pappas}}]{Bannenberg2018}%
  \BibitemOpen
  \bibfield  {author} {\bibinfo {author} {\bibfnamefont {L.~J.}\ \bibnamefont
  {Bannenberg}}, \bibinfo {author} {\bibfnamefont {F.}~\bibnamefont {Weber}},
  \bibinfo {author} {\bibfnamefont {A.~J.~E.}\ \bibnamefont {Lefering}},
  \bibinfo {author} {\bibfnamefont {T.}~\bibnamefont {Wolf}}, \ and\ \bibinfo
  {author} {\bibfnamefont {C.}~\bibnamefont {Pappas}},\ }\bibfield  {title}
  {\enquote {\bibinfo {title} {Magnetization and ac susceptibility study of the
  cubic chiral magnet
  ${\mathrm{mn}}_{1\ensuremath{-}x}{\mathrm{fe}}_{x}\mathrm{Si}$},}\ }\href
  {\doibase 10.1103/PhysRevB.98.184430} {\bibfield  {journal} {\bibinfo
  {journal} {Phys. Rev. B}\ }\textbf {\bibinfo {volume} {98}},\ \bibinfo
  {pages} {184430} (\bibinfo {year} {2018})}\BibitemShut {NoStop}%
\bibitem [{\citenamefont {Vistoli}\ \emph {et~al.}(2019)\citenamefont
  {Vistoli}, \citenamefont {Wang}, \citenamefont {Sander}, \citenamefont {Zhu},
  \citenamefont {Casals}, \citenamefont {Cichelero}, \citenamefont
  {Barth{\'e}{\'e}my}, \citenamefont {Fusil}, \citenamefont {Herranz},
  \citenamefont {Valencia}, \citenamefont {Abrudan}, \citenamefont {Weschke},
  \citenamefont {Nakazawa}, \citenamefont {Kohno}, \citenamefont {Santamaria},
  \citenamefont {Wu}, \citenamefont {Garcia},\ and\ \citenamefont
  {Bibes}}]{Vistoli2019}%
  \BibitemOpen
  \bibfield  {author} {\bibinfo {author} {\bibfnamefont {Lorenzo}\ \bibnamefont
  {Vistoli}}, \bibinfo {author} {\bibfnamefont {Wenbo}\ \bibnamefont {Wang}},
  \bibinfo {author} {\bibfnamefont {Anke}\ \bibnamefont {Sander}}, \bibinfo
  {author} {\bibfnamefont {Qiuxiang}\ \bibnamefont {Zhu}}, \bibinfo {author}
  {\bibfnamefont {Blai}\ \bibnamefont {Casals}}, \bibinfo {author}
  {\bibfnamefont {Rafael}\ \bibnamefont {Cichelero}}, \bibinfo {author}
  {\bibfnamefont {Agn{\`e}s}\ \bibnamefont {Barth{\'e}{\'e}my}}, \bibinfo
  {author} {\bibfnamefont {St{\'e}phane}\ \bibnamefont {Fusil}}, \bibinfo
  {author} {\bibfnamefont {Gervasi}\ \bibnamefont {Herranz}}, \bibinfo {author}
  {\bibfnamefont {Sergio}\ \bibnamefont {Valencia}}, \bibinfo {author}
  {\bibfnamefont {Radu}\ \bibnamefont {Abrudan}}, \bibinfo {author}
  {\bibfnamefont {Eugen}\ \bibnamefont {Weschke}}, \bibinfo {author}
  {\bibfnamefont {Kazuki}\ \bibnamefont {Nakazawa}}, \bibinfo {author}
  {\bibfnamefont {Hiroshi}\ \bibnamefont {Kohno}}, \bibinfo {author}
  {\bibfnamefont {Jacobo}\ \bibnamefont {Santamaria}}, \bibinfo {author}
  {\bibfnamefont {Weida}\ \bibnamefont {Wu}}, \bibinfo {author} {\bibfnamefont
  {Vincent}\ \bibnamefont {Garcia}}, \ and\ \bibinfo {author} {\bibfnamefont
  {Manuel}\ \bibnamefont {Bibes}},\ }\bibfield  {title} {\enquote {\bibinfo
  {title} {Giant topological hall effect in correlated oxide thin films},}\
  }\href@noop {} {\bibfield  {journal} {\bibinfo  {journal} {Nature Physics}\
  }\textbf {\bibinfo {volume} {15}},\ \bibinfo {pages} {67--72} (\bibinfo
  {year} {2019})}\BibitemShut {NoStop}%
\bibitem [{\citenamefont {Nakamura}\ \emph {et~al.}(2018)\citenamefont
  {Nakamura}, \citenamefont {Morikawa}, \citenamefont {Yu}, \citenamefont
  {Kagawa}, \citenamefont {Arima}, \citenamefont {Tokura},\ and\ \citenamefont
  {Kawasak}}]{Nakamura2018}%
  \BibitemOpen
  \bibfield  {author} {\bibinfo {author} {\bibfnamefont {Masao}\ \bibnamefont
  {Nakamura}}, \bibinfo {author} {\bibfnamefont {Daisuke}\ \bibnamefont
  {Morikawa}}, \bibinfo {author} {\bibfnamefont {X~iuzhen}\ \bibnamefont {Yu}},
  \bibinfo {author} {\bibfnamefont {Fumitaka}\ \bibnamefont {Kagawa}}, \bibinfo
  {author} {\bibfnamefont {Takahisa}\ \bibnamefont {Arima}}, \bibinfo {author}
  {\bibfnamefont {Yoshinori}\ \bibnamefont {Tokura}}, \ and\ \bibinfo {author}
  {\bibfnamefont {Masashi}\ \bibnamefont {Kawasak}},\ }\bibfield  {title}
  {\enquote {\bibinfo {title} {Emergence of topological hall effect in
  half-metallic manganite thin films by tuning perpendicular magnetic
  anisotropy},}\ }\href {\doibase 10.7566/JPSJ.87.074704} {\bibfield  {journal}
  {\bibinfo  {journal} {Journal of the Physical Society of Japan}\ }\textbf
  {\bibinfo {volume} {87}},\ \bibinfo {pages} {074704} (\bibinfo {year}
  {2018})}\BibitemShut {NoStop}%
\bibitem [{\citenamefont {Maccariello}\ \emph {et~al.}(2018)\citenamefont
  {Maccariello}, \citenamefont {Legrand}, \citenamefont {Reyren}, \citenamefont
  {Garcia}, \citenamefont {Bouzehouane}, \citenamefont {Collin}, \citenamefont
  {Cros},\ and\ \citenamefont {Fert}}]{Maccariello2018}%
  \BibitemOpen
  \bibfield  {author} {\bibinfo {author} {\bibfnamefont {Davide}\ \bibnamefont
  {Maccariello}}, \bibinfo {author} {\bibfnamefont {William}\ \bibnamefont
  {Legrand}}, \bibinfo {author} {\bibfnamefont {Nicolas}\ \bibnamefont
  {Reyren}}, \bibinfo {author} {\bibfnamefont {Karin}\ \bibnamefont {Garcia}},
  \bibinfo {author} {\bibfnamefont {Karim}\ \bibnamefont {Bouzehouane}},
  \bibinfo {author} {\bibfnamefont {Sophie}\ \bibnamefont {Collin}}, \bibinfo
  {author} {\bibfnamefont {Vincent}\ \bibnamefont {Cros}}, \ and\ \bibinfo
  {author} {\bibfnamefont {Albert}\ \bibnamefont {Fert}},\ }\bibfield  {title}
  {\enquote {\bibinfo {title} {Electrical detection of single magnetic
  skyrmions in metallic multilayers at room temperature},}\ }\href {\doibase
  10.1038/s41565-017-0044-4} {\bibfield  {journal} {\bibinfo  {journal} {Nature
  Nanotechnology}\ }\textbf {\bibinfo {volume} {13}},\ \bibinfo {pages}
  {748--3395} (\bibinfo {year} {2018})}\BibitemShut {NoStop}%
\bibitem [{\citenamefont {Jeong}\ \emph {et~al.}(2015)\citenamefont {Jeong},
  \citenamefont {Yang}, \citenamefont {Yang}, \citenamefont {Ayala-Valenzuela},
  \citenamefont {Wulferding}, \citenamefont {Zhou}, \citenamefont {Goodenough},
  \citenamefont {de~Lozanne}, \citenamefont {Mitchell}, \citenamefont {Leon},
  \citenamefont {Movshovich}, \citenamefont {Jeong}, \citenamefont {Yeom},\
  and\ \citenamefont {Kim}}]{Jeong2015}%
  \BibitemOpen
  \bibfield  {author} {\bibinfo {author} {\bibfnamefont {Juyoung}\ \bibnamefont
  {Jeong}}, \bibinfo {author} {\bibfnamefont {Ilkyu}\ \bibnamefont {Yang}},
  \bibinfo {author} {\bibfnamefont {Jinho}\ \bibnamefont {Yang}}, \bibinfo
  {author} {\bibfnamefont {Oscar~E.}\ \bibnamefont {Ayala-Valenzuela}},
  \bibinfo {author} {\bibfnamefont {Dirk}\ \bibnamefont {Wulferding}}, \bibinfo
  {author} {\bibfnamefont {J.-S.}\ \bibnamefont {Zhou}}, \bibinfo {author}
  {\bibfnamefont {John~B.}\ \bibnamefont {Goodenough}}, \bibinfo {author}
  {\bibfnamefont {Alex}\ \bibnamefont {de~Lozanne}}, \bibinfo {author}
  {\bibfnamefont {J.~F.}\ \bibnamefont {Mitchell}}, \bibinfo {author}
  {\bibfnamefont {Neliza}\ \bibnamefont {Leon}}, \bibinfo {author}
  {\bibfnamefont {Roman}\ \bibnamefont {Movshovich}}, \bibinfo {author}
  {\bibfnamefont {Yoon~Hee}\ \bibnamefont {Jeong}}, \bibinfo {author}
  {\bibfnamefont {Han~Woong}\ \bibnamefont {Yeom}}, \ and\ \bibinfo {author}
  {\bibfnamefont {Jeehoon}\ \bibnamefont {Kim}},\ }\bibfield  {title} {\enquote
  {\bibinfo {title} {Magnetic domain tuning and the emergence of bubble domains
  in the bilayer manganite {L}a$_{2-2x}${S}r$_{1+2x}${M}n$_2${O}$_7$ ($x =
  0.32$)},}\ }\href {\doibase 10.1103/PhysRevB.92.054426} {\bibfield  {journal}
  {\bibinfo  {journal} {Phys. Rev. B}\ }\textbf {\bibinfo {volume} {92}},\
  \bibinfo {pages} {054426} (\bibinfo {year} {2015})}\BibitemShut {NoStop}%
\bibitem [{\citenamefont {Maus}(2019)}]{Maus2019}%
  \BibitemOpen
  \bibfield  {author} {\bibinfo {author} {\bibfnamefont {Mark}\ \bibnamefont
  {Maus}},\ }\href@noop {} {} (\bibinfo {year} {2019}),\ \bibinfo {note}
  {magnetic AC Susceptometry for Characterizing Magnetic Spin Structures,
  Undergraduate Honor Thesis, University of Colorado Boulder}\BibitemShut
  {NoStop}%
\bibitem [{\citenamefont {Yang}\ \emph {et~al.}(2016)\citenamefont {Yang},
  \citenamefont {Yang}, \citenamefont {Kim}, \citenamefont {Shin},
  \citenamefont {Jeong}, \citenamefont {Wulferding}, \citenamefont {Yeom},\
  and\ \citenamefont {Kim}}]{Yang2016}%
  \BibitemOpen
  \bibfield  {author} {\bibinfo {author} {\bibfnamefont {Jinho}\ \bibnamefont
  {Yang}}, \bibinfo {author} {\bibfnamefont {Ilkyu}\ \bibnamefont {Yang}},
  \bibinfo {author} {\bibfnamefont {Yun~Won}\ \bibnamefont {Kim}}, \bibinfo
  {author} {\bibfnamefont {Dongwoo}\ \bibnamefont {Shin}}, \bibinfo {author}
  {\bibfnamefont {Juyoung}\ \bibnamefont {Jeong}}, \bibinfo {author}
  {\bibfnamefont {Dirk}\ \bibnamefont {Wulferding}}, \bibinfo {author}
  {\bibfnamefont {Han~Woong}\ \bibnamefont {Yeom}}, \ and\ \bibinfo {author}
  {\bibfnamefont {Jeehoon}\ \bibnamefont {Kim}},\ }\bibfield  {title} {\enquote
  {\bibinfo {title} {Construction of a 3he magnetic force microscope with a
  vector magnet},}\ }\href {\doibase 10.1063/1.4941959} {\bibfield  {journal}
  {\bibinfo  {journal} {Review of Scientific Instruments}\ }\textbf {\bibinfo
  {volume} {87}},\ \bibinfo {pages} {023704} (\bibinfo {year}
  {2016})}\BibitemShut {NoStop}%
\bibitem [{\citenamefont {Campbell}\ and\ \citenamefont
  {Fert}(1982)}]{Campbell1982}%
  \BibitemOpen
  \bibfield  {author} {\bibinfo {author} {\bibfnamefont {I.A.}\ \bibnamefont
  {Campbell}}\ and\ \bibinfo {author} {\bibfnamefont {A.}~\bibnamefont
  {Fert}},\ }\bibfield  {title} {\enquote {\bibinfo {title} {Chapter 9
  transport properties of ferromagnets},}\ \ }(\bibinfo  {publisher}
  {Elsevier},\ \bibinfo {year} {1982})\ pp.\ \bibinfo {pages}
  {747--804}\BibitemShut {NoStop}%
\bibitem [{\citenamefont {{De Gennes}}\ and\ \citenamefont
  {Friedel}(1958)}]{deGennes1958}%
  \BibitemOpen
  \bibfield  {author} {\bibinfo {author} {\bibfnamefont {P.G.}\ \bibnamefont
  {{De Gennes}}}\ and\ \bibinfo {author} {\bibfnamefont {J.}~\bibnamefont
  {Friedel}},\ }\bibfield  {title} {\enquote {\bibinfo {title} {Anomalies de
  dans r\'{e}sistivit\'{e} certains m\'{e}taux magn\'{i}ques},}\ }\href
  {\doibase https://doi.org/10.1016/0022-3697(58)90196-3} {\bibfield  {journal}
  {\bibinfo  {journal} {Journal of Physics and Chemistry of Solids}\ }\textbf
  {\bibinfo {volume} {4}},\ \bibinfo {pages} {71--77} (\bibinfo {year}
  {1958})}\BibitemShut {NoStop}%
\bibitem [{\citenamefont {Yosida}(1957{\natexlab{b}})}]{Yosida1957}%
  \BibitemOpen
  \bibfield  {author} {\bibinfo {author} {\bibfnamefont {Kei}\ \bibnamefont
  {Yosida}},\ }\bibfield  {title} {\enquote {\bibinfo {title} {Anomalous
  electrical resistivity and magnetoresistance due to an $s$-$d$ interaction in
  cu-mn alloys},}\ }\href {\doibase 10.1103/PhysRev.107.396} {\bibfield
  {journal} {\bibinfo  {journal} {Phys. Rev.}\ }\textbf {\bibinfo {volume}
  {107}},\ \bibinfo {pages} {396--403} (\bibinfo {year}
  {1957}{\natexlab{b}})}\BibitemShut {NoStop}%
\bibitem [{\citenamefont {Fisher}\ and\ \citenamefont
  {Langer}(1968)}]{Fisher1968}%
  \BibitemOpen
  \bibfield  {author} {\bibinfo {author} {\bibfnamefont {Michael~E.}\
  \bibnamefont {Fisher}}\ and\ \bibinfo {author} {\bibfnamefont {J.~S.}\
  \bibnamefont {Langer}},\ }\bibfield  {title} {\enquote {\bibinfo {title}
  {Resistive anomalies at magnetic critical points},}\ }\href {\doibase
  10.1103/PhysRevLett.20.665} {\bibfield  {journal} {\bibinfo  {journal} {Phys.
  Rev. Lett.}\ }\textbf {\bibinfo {volume} {20}},\ \bibinfo {pages} {665--668}
  (\bibinfo {year} {1968})}\BibitemShut {NoStop}%
\bibitem [{\citenamefont {Alexander}\ \emph {et~al.}(1976)\citenamefont
  {Alexander}, \citenamefont {Helman},\ and\ \citenamefont
  {Balberg}}]{Alexander1976}%
  \BibitemOpen
  \bibfield  {author} {\bibinfo {author} {\bibfnamefont {S.}~\bibnamefont
  {Alexander}}, \bibinfo {author} {\bibfnamefont {J.~S.}\ \bibnamefont
  {Helman}}, \ and\ \bibinfo {author} {\bibfnamefont {I.}~\bibnamefont
  {Balberg}},\ }\bibfield  {title} {\enquote {\bibinfo {title} {Critical
  behavior of the electrical resistivity in magnetic systems},}\ }\href
  {\doibase 10.1103/PhysRevB.13.304} {\bibfield  {journal} {\bibinfo  {journal}
  {Phys. Rev. B}\ }\textbf {\bibinfo {volume} {13}},\ \bibinfo {pages}
  {304--315} (\bibinfo {year} {1976})}\BibitemShut {NoStop}%
\bibitem [{\citenamefont {Wang}\ \emph {et~al.}(2021)\citenamefont {Wang},
  \citenamefont {Howells}, \citenamefont {Marshall}, \citenamefont {Taylor},
  \citenamefont {Edmonds}, \citenamefont {Rushforth}, \citenamefont {Campion},\
  and\ \citenamefont {Gallagher}}]{MWang2021}%
  \BibitemOpen
  \bibfield  {author} {\bibinfo {author} {\bibfnamefont {M.}~\bibnamefont
  {Wang}}, \bibinfo {author} {\bibfnamefont {B.}~\bibnamefont {Howells}},
  \bibinfo {author} {\bibfnamefont {R.~A.}\ \bibnamefont {Marshall}}, \bibinfo
  {author} {\bibfnamefont {J.~M.}\ \bibnamefont {Taylor}}, \bibinfo {author}
  {\bibfnamefont {K.~W.}\ \bibnamefont {Edmonds}}, \bibinfo {author}
  {\bibfnamefont {A.~W.}\ \bibnamefont {Rushforth}}, \bibinfo {author}
  {\bibfnamefont {R.~P.}\ \bibnamefont {Campion}}, \ and\ \bibinfo {author}
  {\bibfnamefont {B.~L.}\ \bibnamefont {Gallagher}},\ }\bibfield  {title}
  {\enquote {\bibinfo {title} {Magnetism and magnetoresistance in the critical
  region of a dilute ferromagnet},}\ }\href {\doibase 10.1103/PhysRev.107.396}
  {\bibfield  {journal} {\bibinfo  {journal} {Scientific Reports}\ }\textbf
  {\bibinfo {volume} {11}},\ \bibinfo {pages} {2300} (\bibinfo {year}
  {2021})}\BibitemShut {NoStop}%
\bibitem [{\citenamefont {Chakravorty}\ and\ \citenamefont
  {Raychaudhuri}(2015)}]{Chakravorty2015}%
  \BibitemOpen
  \bibfield  {author} {\bibinfo {author} {\bibfnamefont {Manotosh}\
  \bibnamefont {Chakravorty}}\ and\ \bibinfo {author} {\bibfnamefont {A.~K.}\
  \bibnamefont {Raychaudhuri}},\ }\bibfield  {title} {\enquote {\bibinfo
  {title} {Magnetoresistance of polycrystalline gadolinium with varying grain
  size},}\ }\href {\doibase 10.1063/1.4904919} {\bibfield  {journal} {\bibinfo
  {journal} {Journal of Applied Physics}\ }\textbf {\bibinfo {volume} {117}},\
  \bibinfo {pages} {034301} (\bibinfo {year} {2015})}\BibitemShut {NoStop}%
\bibitem [{\citenamefont {Checkelsky}\ \emph {et~al.}(2008)\citenamefont
  {Checkelsky}, \citenamefont {Lee}, \citenamefont {Morosan}, \citenamefont
  {Cava},\ and\ \citenamefont {Ong}}]{Checkelsky2008}%
  \BibitemOpen
  \bibfield  {author} {\bibinfo {author} {\bibfnamefont {J.~G.}\ \bibnamefont
  {Checkelsky}}, \bibinfo {author} {\bibfnamefont {Minhyea}\ \bibnamefont
  {Lee}}, \bibinfo {author} {\bibfnamefont {E.}~\bibnamefont {Morosan}},
  \bibinfo {author} {\bibfnamefont {R.~J.}\ \bibnamefont {Cava}}, \ and\
  \bibinfo {author} {\bibfnamefont {N.~P.}\ \bibnamefont {Ong}},\ }\bibfield
  {title} {\enquote {\bibinfo {title} {Anomalous hall effect and
  magnetoresistance in the layered ferromagnet
  ${\mathrm{fe}}_{1/4}\mathrm{Ta}{\mathrm{s}}_{2}$: The inelastic regime},}\
  }\href@noop {} {\bibfield  {journal} {\bibinfo  {journal} {Phys. Rev. B}\
  }\textbf {\bibinfo {volume} {77}},\ \bibinfo {pages} {014433} (\bibinfo
  {year} {2008})}\BibitemShut {NoStop}%
\bibitem [{\citenamefont {Lee}\ \emph {et~al.}(2007)\citenamefont {Lee},
  \citenamefont {Onose}, \citenamefont {Tokura},\ and\ \citenamefont
  {Ong}}]{Lee2007}%
  \BibitemOpen
  \bibfield  {author} {\bibinfo {author} {\bibfnamefont {Minhyea}\ \bibnamefont
  {Lee}}, \bibinfo {author} {\bibfnamefont {Y.}~\bibnamefont {Onose}}, \bibinfo
  {author} {\bibfnamefont {Y.}~\bibnamefont {Tokura}}, \ and\ \bibinfo {author}
  {\bibfnamefont {N.~P.}\ \bibnamefont {Ong}},\ }\bibfield  {title} {\enquote
  {\bibinfo {title} {Hidden constant in the anomalous hall effect of
  high-purity magnet mnsi},}\ }\href {\doibase 10.1103/PhysRevB.75.172403}
  {\bibfield  {journal} {\bibinfo  {journal} {Phys. Rev. B}\ }\textbf {\bibinfo
  {volume} {75}},\ \bibinfo {pages} {172403} (\bibinfo {year}
  {2007})}\BibitemShut {NoStop}%
\bibitem [{\citenamefont {Huang}\ and\ \citenamefont
  {Chien}(2012)}]{Huang2012}%
  \BibitemOpen
  \bibfield  {author} {\bibinfo {author} {\bibfnamefont {SX}~\bibnamefont
  {Huang}}\ and\ \bibinfo {author} {\bibfnamefont {CL}~\bibnamefont {Chien}},\
  }\bibfield  {title} {\enquote {\bibinfo {title} {Extended skyrmion phase in
  epitaxial fege (111) thin films},}\ }\href@noop {} {\bibfield  {journal}
  {\bibinfo  {journal} {Physical review letters}\ }\textbf {\bibinfo {volume}
  {108}},\ \bibinfo {pages} {267201} (\bibinfo {year} {2012})}\BibitemShut
  {NoStop}%
\bibitem [{\citenamefont {Chapman}\ \emph {et~al.}(2013)\citenamefont
  {Chapman}, \citenamefont {Grossnickle}, \citenamefont {Wolf},\ and\
  \citenamefont {Lee}}]{Chapman2013}%
  \BibitemOpen
  \bibfield  {author} {\bibinfo {author} {\bibfnamefont {Benjamin~J.}\
  \bibnamefont {Chapman}}, \bibinfo {author} {\bibfnamefont {Maxwell~G.}\
  \bibnamefont {Grossnickle}}, \bibinfo {author} {\bibfnamefont {Thomas}\
  \bibnamefont {Wolf}}, \ and\ \bibinfo {author} {\bibfnamefont {Minhyea}\
  \bibnamefont {Lee}},\ }\bibfield  {title} {\enquote {\bibinfo {title} {Large
  enhancement of emergent magnetic fields in mnsi with impurities and
  pressure},}\ }\href {\doibase 10.1103/PhysRevB.88.214406} {\bibfield
  {journal} {\bibinfo  {journal} {Phys. Rev. B}\ }\textbf {\bibinfo {volume}
  {88}},\ \bibinfo {pages} {214406} (\bibinfo {year} {2013})}\BibitemShut
  {NoStop}%
\bibitem [{\citenamefont {Nagaosa}\ \emph {et~al.}(2010)\citenamefont
  {Nagaosa}, \citenamefont {Sinova}, \citenamefont {Onoda}, \citenamefont
  {MacDonald},\ and\ \citenamefont {Ong}}]{AHEreview2010}%
  \BibitemOpen
  \bibfield  {author} {\bibinfo {author} {\bibfnamefont {Naoto}\ \bibnamefont
  {Nagaosa}}, \bibinfo {author} {\bibfnamefont {Jairo}\ \bibnamefont {Sinova}},
  \bibinfo {author} {\bibfnamefont {Shigeki}\ \bibnamefont {Onoda}}, \bibinfo
  {author} {\bibfnamefont {A.~H.}\ \bibnamefont {MacDonald}}, \ and\ \bibinfo
  {author} {\bibfnamefont {N.~P.}\ \bibnamefont {Ong}},\ }\bibfield  {title}
  {\enquote {\bibinfo {title} {Anomalous hall effect},}\ }\href {\doibase
  10.1103/RevModPhys.82.1539} {\bibfield  {journal} {\bibinfo  {journal} {Rev.
  Mod. Phys.}\ }\textbf {\bibinfo {volume} {82}},\ \bibinfo {pages}
  {1539--1592} (\bibinfo {year} {2010})}\BibitemShut {NoStop}%
\bibitem [{\citenamefont {Kanazawa}\ \emph {et~al.}(2011)\citenamefont
  {Kanazawa}, \citenamefont {Onose}, \citenamefont {Arima}, \citenamefont
  {Okuyama}, \citenamefont {Ohoyama}, \citenamefont {Wakimoto}, \citenamefont
  {Kakurai}, \citenamefont {Ishiwata},\ and\ \citenamefont
  {Tokura}}]{Kanazawa2011}%
  \BibitemOpen
  \bibfield  {author} {\bibinfo {author} {\bibfnamefont {N.}~\bibnamefont
  {Kanazawa}}, \bibinfo {author} {\bibfnamefont {Y.}~\bibnamefont {Onose}},
  \bibinfo {author} {\bibfnamefont {T.}~\bibnamefont {Arima}}, \bibinfo
  {author} {\bibfnamefont {D.}~\bibnamefont {Okuyama}}, \bibinfo {author}
  {\bibfnamefont {K.}~\bibnamefont {Ohoyama}}, \bibinfo {author} {\bibfnamefont
  {S.}~\bibnamefont {Wakimoto}}, \bibinfo {author} {\bibfnamefont
  {K.}~\bibnamefont {Kakurai}}, \bibinfo {author} {\bibfnamefont
  {S.}~\bibnamefont {Ishiwata}}, \ and\ \bibinfo {author} {\bibfnamefont
  {Y.}~\bibnamefont {Tokura}},\ }\bibfield  {title} {\enquote {\bibinfo {title}
  {Large topological hall effect in a short-period helimagnet mnge},}\ }\href
  {\doibase 10.1103/PhysRevLett.106.156603} {\bibfield  {journal} {\bibinfo
  {journal} {Phys. Rev. Lett.}\ }\textbf {\bibinfo {volume} {106}},\ \bibinfo
  {pages} {156603} (\bibinfo {year} {2011})}\BibitemShut {NoStop}%
\bibitem [{\citenamefont {Shiomi}\ \emph {et~al.}(2012)\citenamefont {Shiomi},
  \citenamefont {Iguchi},\ and\ \citenamefont {Tokura}}]{Shiomi2012}%
  \BibitemOpen
  \bibfield  {author} {\bibinfo {author} {\bibfnamefont {Y.}~\bibnamefont
  {Shiomi}}, \bibinfo {author} {\bibfnamefont {S.}~\bibnamefont {Iguchi}}, \
  and\ \bibinfo {author} {\bibfnamefont {Y.}~\bibnamefont {Tokura}},\
  }\bibfield  {title} {\enquote {\bibinfo {title} {Emergence of topological
  hall effect from fanlike spin structure as modified by dzyaloshinsky-moriya
  interaction in mnp},}\ }\href {\doibase 10.1103/PhysRevB.86.180404}
  {\bibfield  {journal} {\bibinfo  {journal} {Phys. Rev. B}\ }\textbf {\bibinfo
  {volume} {86}},\ \bibinfo {pages} {180404} (\bibinfo {year}
  {2012})}\BibitemShut {NoStop}%
\bibitem [{\citenamefont {Franz}\ \emph {et~al.}(2014)\citenamefont {Franz},
  \citenamefont {Freimuth}, \citenamefont {Bauer}, \citenamefont {Ritz},
  \citenamefont {Schnarr}, \citenamefont {Duvinage}, \citenamefont {Adams},
  \citenamefont {Bl\"ugel}, \citenamefont {Rosch}, \citenamefont {Mokrousov},\
  and\ \citenamefont {Pfleiderer}}]{Franz2014}%
  \BibitemOpen
  \bibfield  {author} {\bibinfo {author} {\bibfnamefont {C.}~\bibnamefont
  {Franz}}, \bibinfo {author} {\bibfnamefont {F.}~\bibnamefont {Freimuth}},
  \bibinfo {author} {\bibfnamefont {A.}~\bibnamefont {Bauer}}, \bibinfo
  {author} {\bibfnamefont {R.}~\bibnamefont {Ritz}}, \bibinfo {author}
  {\bibfnamefont {C.}~\bibnamefont {Schnarr}}, \bibinfo {author} {\bibfnamefont
  {C.}~\bibnamefont {Duvinage}}, \bibinfo {author} {\bibfnamefont
  {T.}~\bibnamefont {Adams}}, \bibinfo {author} {\bibfnamefont
  {S.}~\bibnamefont {Bl\"ugel}}, \bibinfo {author} {\bibfnamefont
  {A.}~\bibnamefont {Rosch}}, \bibinfo {author} {\bibfnamefont
  {Y.}~\bibnamefont {Mokrousov}}, \ and\ \bibinfo {author} {\bibfnamefont
  {C.}~\bibnamefont {Pfleiderer}},\ }\bibfield  {title} {\enquote {\bibinfo
  {title} {Real-space and reciprocal-space berry phases in the hall effect of
  ${\mathrm{mn}}_{1\ensuremath{-}x}{\mathrm{fe}}_{x}\mathrm{Si}$},}\ }\href
  {\doibase 10.1103/PhysRevLett.112.186601} {\bibfield  {journal} {\bibinfo
  {journal} {Phys. Rev. Lett.}\ }\textbf {\bibinfo {volume} {112}},\ \bibinfo
  {pages} {186601} (\bibinfo {year} {2014})}\BibitemShut {NoStop}%
\bibitem [{\citenamefont {Bauer}\ \emph {et~al.}(2010)\citenamefont {Bauer},
  \citenamefont {Neubauer}, \citenamefont {Franz}, \citenamefont {M\"unzer},
  \citenamefont {Garst},\ and\ \citenamefont {Pfleiderer}}]{Bauer2010}%
  \BibitemOpen
  \bibfield  {author} {\bibinfo {author} {\bibfnamefont {A.}~\bibnamefont
  {Bauer}}, \bibinfo {author} {\bibfnamefont {A.}~\bibnamefont {Neubauer}},
  \bibinfo {author} {\bibfnamefont {C.}~\bibnamefont {Franz}}, \bibinfo
  {author} {\bibfnamefont {W.}~\bibnamefont {M\"unzer}}, \bibinfo {author}
  {\bibfnamefont {M.}~\bibnamefont {Garst}}, \ and\ \bibinfo {author}
  {\bibfnamefont {C.}~\bibnamefont {Pfleiderer}},\ }\bibfield  {title}
  {\enquote {\bibinfo {title} {Quantum phase transitions in single-crystal
  ${\text{mn}}_{1\ensuremath{-}x}{\text{fe}}_{x}\text{Si}$ and
  ${\text{mn}}_{1\ensuremath{-}x}{\text{co}}_{x}\text{Si}$: Crystal growth,
  magnetization, ac susceptibility, and specific heat},}\ }\href {\doibase
  10.1103/PhysRevB.82.064404} {\bibfield  {journal} {\bibinfo  {journal} {Phys.
  Rev. B}\ }\textbf {\bibinfo {volume} {82}},\ \bibinfo {pages} {064404}
  (\bibinfo {year} {2010})}\BibitemShut {NoStop}%
\bibitem [{afn()}]{afnote}%
  \BibitemOpen
  \href@noop {} {}\bibinfo {note} {The area fraction is determined by dividing
  the centered MFM images into negative and positive force values, and
  subsequently calculating the ratio of the respective areas.}\BibitemShut
  {Stop}%
\bibitem [{\citenamefont {Shinjo}\ \emph {et~al.}(2000)\citenamefont {Shinjo},
  \citenamefont {Okuno}, \citenamefont {Hassdorf}, \citenamefont {Shigeto},\
  and\ \citenamefont {Ono}}]{Shinjo2000}%
  \BibitemOpen
  \bibfield  {author} {\bibinfo {author} {\bibfnamefont {T.}~\bibnamefont
  {Shinjo}}, \bibinfo {author} {\bibfnamefont {T.}~\bibnamefont {Okuno}},
  \bibinfo {author} {\bibfnamefont {R.}~\bibnamefont {Hassdorf}}, \bibinfo
  {author} {\bibfnamefont {{\textdagger}~K.}\ \bibnamefont {Shigeto}}, \ and\
  \bibinfo {author} {\bibfnamefont {T.}~\bibnamefont {Ono}},\ }\bibfield
  {title} {\enquote {\bibinfo {title} {Magnetic vortex core observation in
  circular dots of permalloy},}\ }\href {\doibase 10.1126/science.289.5481.930}
  {\bibfield  {journal} {\bibinfo  {journal} {Science}\ }\textbf {\bibinfo
  {volume} {289}},\ \bibinfo {pages} {930--932} (\bibinfo {year}
  {2000})}\BibitemShut {NoStop}%
\bibitem [{\citenamefont {Wachowiak}\ \emph {et~al.}(2002)\citenamefont
  {Wachowiak}, \citenamefont {Wiebe}, \citenamefont {Bode}, \citenamefont
  {Pietzsch}, \citenamefont {Morgenstern},\ and\ \citenamefont
  {Wiesendanger}}]{Wachowiak2002}%
  \BibitemOpen
  \bibfield  {author} {\bibinfo {author} {\bibfnamefont {A.}~\bibnamefont
  {Wachowiak}}, \bibinfo {author} {\bibfnamefont {J.}~\bibnamefont {Wiebe}},
  \bibinfo {author} {\bibfnamefont {M.}~\bibnamefont {Bode}}, \bibinfo {author}
  {\bibfnamefont {O.}~\bibnamefont {Pietzsch}}, \bibinfo {author}
  {\bibfnamefont {M.}~\bibnamefont {Morgenstern}}, \ and\ \bibinfo {author}
  {\bibfnamefont {R.}~\bibnamefont {Wiesendanger}},\ }\bibfield  {title}
  {\enquote {\bibinfo {title} {Direct observation of internal spin structure of
  magnetic vortex cores},}\ }\href {\doibase 10.1126/science.1075302}
  {\bibfield  {journal} {\bibinfo  {journal} {Science}\ }\textbf {\bibinfo
  {volume} {298}},\ \bibinfo {pages} {577--580} (\bibinfo {year}
  {2002})}\BibitemShut {NoStop}%
\bibitem [{\citenamefont {Niitsu}\ \emph {et~al.}(2018)\citenamefont {Niitsu},
  \citenamefont {Tanigaki}, \citenamefont {Harada},\ and\ \citenamefont
  {Shindo}}]{Niitsu2018}%
  \BibitemOpen
  \bibfield  {author} {\bibinfo {author} {\bibfnamefont {K.}~\bibnamefont
  {Niitsu}}, \bibinfo {author} {\bibfnamefont {T.}~\bibnamefont {Tanigaki}},
  \bibinfo {author} {\bibfnamefont {K.}~\bibnamefont {Harada}}, \ and\ \bibinfo
  {author} {\bibfnamefont {D.}~\bibnamefont {Shindo}},\ }\bibfield  {title}
  {\enquote {\bibinfo {title} {Temperature dependence of 180$^{\circ}$ domain
  wall width in iron and nickel films analyzed using electron holography},}\
  }\href {\doibase 10.1063/1.5056308} {\bibfield  {journal} {\bibinfo
  {journal} {Applied Physics Letters}\ }\textbf {\bibinfo {volume} {113}},\
  \bibinfo {pages} {222407} (\bibinfo {year} {2018})}\BibitemShut {NoStop}%
\bibitem [{\citenamefont {Raabe}\ \emph {et~al.}(2000)\citenamefont {Raabe},
  \citenamefont {Pulwey}, \citenamefont {Sattler}, \citenamefont {Schweinbk},
  \citenamefont {Zweck},\ and\ \citenamefont {Weiss}}]{Raabe2000}%
  \BibitemOpen
  \bibfield  {author} {\bibinfo {author} {\bibfnamefont {J.}~\bibnamefont
  {Raabe}}, \bibinfo {author} {\bibfnamefont {R.}~\bibnamefont {Pulwey}},
  \bibinfo {author} {\bibfnamefont {R.}~\bibnamefont {Sattler}}, \bibinfo
  {author} {\bibfnamefont {T.}~\bibnamefont {Schweinbk}}, \bibinfo {author}
  {\bibfnamefont {J.}~\bibnamefont {Zweck}}, \ and\ \bibinfo {author}
  {\bibfnamefont {D.}~\bibnamefont {Weiss}},\ }\bibfield  {title} {\enquote
  {\bibinfo {title} {Magnetization pattern of ferromagnetic nanodisks},}\
  }\href {\doibase 10.1063/1.1289216} {\bibfield  {journal} {\bibinfo
  {journal} {Journal of Applied Physics}\ }\textbf {\bibinfo {volume} {88}},\
  \bibinfo {pages} {4437--4439} (\bibinfo {year} {2000})}\BibitemShut {NoStop}%
\bibitem [{\citenamefont {Mesler}\ \emph {et~al.}(2012)\citenamefont {Mesler},
  \citenamefont {Buchanan}, \citenamefont {Im},\ and\ \citenamefont
  {Fischer}}]{Mesler2012}%
  \BibitemOpen
  \bibfield  {author} {\bibinfo {author} {\bibfnamefont {Brooke~L.}\
  \bibnamefont {Mesler}}, \bibinfo {author} {\bibfnamefont {Kristen~S.}\
  \bibnamefont {Buchanan}}, \bibinfo {author} {\bibfnamefont {Mi-Young}\
  \bibnamefont {Im}}, \ and\ \bibinfo {author} {\bibfnamefont {Peter}\
  \bibnamefont {Fischer}},\ }\bibfield  {title} {\enquote {\bibinfo {title}
  {X-ray imaging of nonlinear resonant gyrotropic magnetic vortex core motion
  in circular permalloy disks},}\ }\href {\doibase 10.1063/1.3678448}
  {\bibfield  {journal} {\bibinfo  {journal} {Journal of Applied Physics}\
  }\textbf {\bibinfo {volume} {111}},\ \bibinfo {pages} {07D311} (\bibinfo
  {year} {2012})}\BibitemShut {NoStop}%
\end{thebibliography}

%

\end{document}